\documentclass[11pt,a4paper]{article}
\pdfoutput=1
\usepackage{jheppub}
\usepackage{slashed}

\makeatletter
\def\@fpheader{\relax}
\makeatother

\usepackage[czech,english]{babel}
\usepackage{graphicx}
\usepackage{amsmath,amsfonts,amssymb}
\DeclareMathOperator{\MyProd}{\scalebox{1.4}{$\mathrm{I\kern-0.2ex I}$}}

\preprint{YITP-SB-17-17}

\title{Entanglement Entropy of ABJM Theory and Entropy of Topological Black Hole}

\author[a]{Jun Nian}

\emailAdd{nian@ihes.fr}

\author[b]{Xinyu Zhang}

\emailAdd{zhangxinyuphysics@gmail.com}

\affiliation[a]{Institut des Hautes \'Etudes Scientifiques\\
	Le Bois-Marie, 35 route de Chartres\\
         91440 Bures-sur-Yvette, France\\}

\affiliation[b]{C.N. Yang Institute for Theoretical Physics\\
                    Stony Brook University\\
                    Stony Brook, NY 11794-3840, USA\\}

\abstract{In this paper we discuss the supersymmetric localization of the 4D $\mathcal{N}=2$ off-shell gauged supergravity in the background of the $\textrm{AdS}_4$ neutral topological black hole, which is the gravity dual of the ABJM theory defined on the boundary $\textrm{S}^1 \times \mathbb{H}^2$. We compute the large-$N$ expansion of the supergravity partition function. The result gives the black hole entropy with the logarithmic correction, which matches the previous result of the entanglement entropy of the ABJM theory up to some stringy effects. Our result 
is consistent with the previous on-shell one-loop computation of the logarithmic correction to black hole entropy. It
provides an explicit example of the identification of the entanglement entropy of the boundary conformal field theory with the bulk black hole entropy beyond the leading order given by the classical Bekenstein-Hawking formula, which consequently tests the AdS/CFT correspondence at the subleading order.}


\arxivnumber{}

\newcommand{\bea}{\begin{eqnarray}}
\newcommand{\eea}{\end{eqnarray}}

\newcommand{\be}{\begin{equation}}
\newcommand{\ee}{\end{equation}}

\begin{document}
\maketitle

%%%%%%%%%%%%%%%%%%%%%%%%%%%%%%%%%%%%%%%%%%%
%%%%%%%%%%%%%%%%%%%%%%%%%%%%%%%%%%%%%%%%%%%
\section{Introduction}\label{sec:introduction}
%%%%%%%%%%%%%%%%%%%%%%%%%%%%%%%%%%%%%%%%%%%
%%%%%%%%%%%%%%%%%%%%%%%%%%%%%%%%%%%%%%%%%%%

The interpretation of the black hole entropy is one of the central problems in theoretical physics. The celebrated AdS/CFT correspondence \cite{Maldacena} provides us with a new insight into the problem of the black hole entropy. Based on this principle, the conformal field theory defined on the boundary of an AdS space should capture all features of the gravity theory in the bulk. Hence, it is really tempting to identify the black hole entropy in the bulk and the entanglement entropy of the conformal field theory on the boundary \cite{RT}. When the boundary conformal field theory and the bulk gravity both have certain amount of supersymmetries, the technique of supersymmetric localization allows to compute the entropy on both sides and test the identification precisely. In this paper, we would like to study a concrete example towards this direction, i.e. the ABJM theory via the supergravity localization, to test this proposal.

As a generalization of entanglement entropy, supersymmetric R\'enyi entropy $\textrm{S}_q$ was first defined on a $q$-branched three-sphere \cite{Nishioka-1}. It can be computed exactly using the technique of supersymmetric localization, and the result can be expressed in terms of the partition function of the 3D superconformal field theory defined on a squashed sphere $\textrm{S}_b^3$ with the squashing parameter $b$ \cite{Nishioka-1}. Using the technique of supersymmetric localization on curved manifolds \cite{Pestun}, one can further express this partition function into a matrix integral \cite{HHL, IY, 3D, Nian, Tanaka}. In some cases, one can even evaluate the matrix integral to obtain a relatively simple result. For instance, neglecting the nonperturbative effects at large $N$, the matrix integral for the ABJM theory on some compact manifolds can be evaluated using the Fermi gas approach \cite{Marino, Hatsuda}.

Since it can be computed exactly, the supersymmetric R\'enyi entropy provides a new quantitiy to test various dualities.  For instance, one can use it to test the AdS/CFT correspondence more precisely. Before doing it, one has to first find the holographic way of computing the supersymmetric R\'enyi entropy. As explained in Ref.~\cite{Nishioka-1}, the technical problem is the conical singularity caused by the branched sphere. To resolve the conical singularity, it was proposed in Refs.~\cite{Rey, Nishioka-2} to perform a conformal transformation, which maps the branched three-sphere into $\textrm{S}^1 \times \mathbb{H}^2$, i.e.,
\begin{align}
  ds^2 & = d\theta^2 + q^2 \, \textrm{sin}^2 \theta\, d\tau^2 + \textrm{cos}^2 \theta\, d\phi^2 \nonumber\\
  {} & = \textrm{sin}^2 \theta \left[d\tilde{\tau}^2 + du^2 + \textrm{sinh}^2 u \, d\phi^2 \right]\, ,
\end{align}
where $\tau \in [0,\, 2 \pi)$, $\tilde{\tau} = q \tau \in [0,\, 2 q \pi)$, and
\be
  \textrm{sinh}\, u = - \textrm{cot}\, \theta\, .
\ee
After the conformal transformation, one finds that $\textrm{S}^1 \times \mathbb{H}^2$ can be viewed as the boundary of the $\textrm{AdS}_4$ topological black hole (TBH). Hence, in principle the supersymmetric R\'enyi entropy can be computed in the topological black hole holographically. The free energy and the Killing spinor equations can also be evaluated in the bulk gravity theory, which supports the holographic interpretation \cite{Rey, Nishioka-2}. In particular, the parameter $q$ coming from the branched sphere can be viewed as a deformation parameter of the original theory, related to the mass and the charge of the topological black hole. This relation was called the TBH/qSCFT correspondence \cite{Rey, 4D5D-1}, where it provides another precise test of the AdS/CFT correspondence. Later, these works were generalized to other dimensions, and similar results were found \cite{4D5D-1, 4D5D-2, 5D6D-1, 5D6D-2, 2D3D-1, 2D3D-2, Yang-1, Yang-2}.

The partition functions and consequently the supersymmetric R\'enyi entropies of some superconformal field theories can be computed exactly using the technique of supersymmetric localization. In fact, this technique can also be applied to some supergravity theories. Different backgrounds ($\textrm{AdS}_2 \times \textrm{S}^2$, $\textrm{AdS}_4$) have been studied \cite{BH2010, BH2011, 4DSUGRAloc}. In particular, the localization of the 4D $\mathcal{N}=2$ off-shell supergravity on $\textrm{AdS}_4$ corresponds to the ABJM theory on the boundary $\textrm{S}^3$ \cite{4DSUGRAloc}, and the partition functions of both theories can be expressed in terms of Airy function. From the partition function, we can compute the entanglement entropy of the ABJM theory across a circle $\textrm{S}^1$ on the boundary, which matches the previous results \cite{Nishioka-1}.

It is then natural to consider the supergravity localization on 4D topological black holes, whose boundaries are $\textrm{S}^1 \times \mathbb{H}^2$. From the supergravity localization we should be able to compute holographically the supersymmetric R\'enyi entropies of the corresponding superconformal field theories on the boundary. Comparing the results from the bulk and the known results from the boundary provides an exact test of the AdS/CFT correspondence, and at the same time one can also check the proposal of identifying these entropies as the black hole entropies in this framework concretely.

As a starting point, in this paper we study the localization of the 4D $\mathcal{N}=2$ off-shell gauged supergravity on the neutral topological black hole, which corresponds to the entanglement entropy of the superconformal ABJM theory across a circle $\textrm{S}^1$ on the boundary.
The logic of our computation is as follows. The gravity dual of the ABJM theory is the 11-dimensional M-theory on $\textrm{AdS}_4\times \textrm{S}^7 / \mathbb{Z}_k$ \cite{ABJM}. We neglect all the stringy effects and consistently truncate the 11-dimensional supergravity to a 4-dimensional $\mathcal{N}=2$ gauged supergravity theory, which has an off-shell formalism using superconformal gauged supergravity. We fix the values of fields in the Weyl multiplet, and apply the localization method to evaluate the supersymmetric partition function by integrating over the vector multiplets and the hypermultiplets. Our localization calculation is similar to the standard field theory localization, except that the background spacetime is noncompact. We find that the entropy of the $\textrm{AdS}_4$ neutral topological black hole and the entanglement entropy of the ABJM theory on the boundary coincide in the large-$N$ expansion up to some stringy effects. More precisely,
\be
   S_{EE}^{\textrm{ABJM}} = S_{BH} = - \frac{\sqrt{2} \pi}{3} k^{1/2} N^{3/2} - \frac{1}{4} \textrm{log}(N) + \mathcal{O} (N^0) \, .
\ee
Meanwhile, using the supergravity localization we obtain the logarithmic correction to the classical result of the black hole entropy given by the Bekenstein-Hawking formula \cite{Hawking, Bekenstein-1, Bekenstein-2}, and this correction is consistent with the on-shell 1-loop computation from the Euclidean 11-dimensional supergravity on $\textrm{AdS}_4 \times X_7$ \cite{SenMarino}.

This paper is organized as follows. In Section~\ref{sec:SRE} we review some facts about the supersymmetric R\'enyi entropy and the ABJM theory. The gravity dual of the supersymmetric R\'enyi entropy will be reviewed in Section~\ref{sec:GravDual}. In Section~\ref{sec:SUGRAloc} we discuss the localization of the 4D $\mathcal{N}=2$ off-shell supergravity on the $\textrm{AdS}_4$ neutral topological black hole with the boundary $\textrm{S}^1 \times \mathbb{H}^2$. The bulk black hole entropy and the boundary entanglement entropy of the ABJM theory can be read off from the results of the supergravity localization, which is presented in Section~\ref{sec:Entropy}. Some further discussions will be made in Section~\ref{sec:discussion}. We also present some details of the calculations in a few appendices. In Appendix~\ref{app:SUGRA} we review the 4D $\mathcal{N}=2$ off-shell supergravity, while in Appendix~\ref{app:Killing} the Killing spinors and the convenction of the Gamma matrices are discussed. For the supergravity localization, the explicit form of the localization action will be presented in Appendix~\ref{app:LocAction}, and we will evaluate the action along the localization locus in Appendix~\ref{app:Action}.

%%%%%%%%%%%%%%%%%%%%%%%%%%%%%%%%%%%%%%%%%%%
%%%%%%%%%%%%%%%%%%%%%%%%%%%%%%%%%%%%%%%%%%%
\section{Supersymmetric R\'enyi Entropy of ABJM Theory}\label{sec:SRE}
%%%%%%%%%%%%%%%%%%%%%%%%%%%%%%%%%%%%%%%%%%%
%%%%%%%%%%%%%%%%%%%%%%%%%%%%%%%%%%%%%%%%%%%

\subsection{Supersymmetric R\'enyi Entropy}
We start with the well-known definitions of entanglement entropy and R\'enyi entropy. Suppose the space on which the theory is defined can be divided into a piece $A$ and its complement $\bar{A}=B$, and correspondingly the Hilbert space factorizes into a tensor product $\mathcal{H}_A \otimes \mathcal{H}_B$. If the density matrix over the whole Hilbert space is $\rho$, then the reduced density matrix is defined as
\be
  \rho_A \equiv \textrm{tr}_B \rho\, .
\ee
The entanglement entropy is the von Neumann entropy of $\rho_A$,
\be
  S_E \equiv - \textrm{tr} \rho_A \, \textrm{log} \, \rho_A\, ,
\ee
while the R\'enyi entropies are defined to be 
%To calculate the entanglement entropy, one usually first calculates a related quantity, 
%which is called the R\'enyi entropy. Its definition is
\be\label{eq:DefRenyi}
  S_n \equiv \frac{1}{1-n} \textrm{log}\, \textrm{tr} (\rho_A)^n\, .
\ee
Assuming an analytic continuation of $S_n$ can be obtained, the entanglement entropy can alternately be expressed 
as a limit of the R\'enyi entropy:
%The relation between the R\'enyi entropy and the entanglement entropy is
\be
  \lim_{n\to 1} S_n = S_E\, .
  \label{analyticcont}
\ee
The R\'enyi entropy can be calculated using the so-called ``replica trick'':
\be
  S_n = \frac{1}{1-n} \textrm{log} \left(\frac{Z_n}{(Z_1)^n} \right)\, ,
\ee
where $Z_n$ is the Euclidean partition function on a $n$-covering space branched along $A$.

The concept of the supersymmetric R\'enyi entropy is a generalization of R\'enyi entropy. It was first introduced in Ref.~\cite{Nishioka-1} for the 3-dimensional supersymmetric field theories as follows:
\be
  S_q^{\textrm{SUSY}} \equiv \frac{1}{1-q} \left[\textrm{log} \left(\frac{Z_{\textrm{singular space}} (q)}{(Z_{\textrm{S}^3})^q} \right) \right]\, ,
\ee
where $Z_{\textrm{S}^3}$ is the partition function of a supersymmetric theory on a three-sphere $\textrm{S}^3$, while $Z_{\textrm{singular space}} (q)$ is the partition function on the q-covering of a three-sphere, $\textrm{S}^3_q$,  which is also called the q-branched sphere given by the metric
\be
  ds^2 = L^2 (d\theta^2 + q^2 \, \textrm{sin}^2 \theta\, d\tau^2 + \textrm{cos}^2 \theta\, d\phi^2)
\ee
with $\theta \in [0, \pi/2]$, $\tau \in [0, 2 \pi)$ and $\phi \in [0, 2 \pi)$. In the limit $q \to 1$, the $q$-branched sphere returns to the round sphere, and the supersymmetric R\'enyi entropy becomes the entanglement entropy. Initially, the supersymmetric R\'enyi entropy was defined for 3D $\mathcal{N}=2$ superconformal field theories \cite{Nishioka-1}, and later it was generalized to other dimensions \cite{4D5D-1, 4D5D-2, 5D6D-1, 5D6D-2, 2D3D-1, 2D3D-2, Yang-1, Yang-2}.

Using the supersymmetric localization, it was derived explicitly in Ref.~\cite{Nishioka-1} that in the definition of the supersymmetric R\'enyi entropy of 3D superconformal field theories $Z_{\textrm{singular space}} (q)$ can be written as:
\be
  Z_{\textrm{singular space}} (q) = \frac{1}{|W|} \int \prod_{i=1}^{\textrm{rank}\, G} d\sigma_i\, e^{\pi i k \textrm{Tr} (\sigma^2)} \, \prod_\alpha \frac{1}{\Gamma_h (\alpha(\sigma))}\, \prod_I \prod_{\rho \in \mathcal{R}_I} \Gamma_I \left(\rho(\sigma) + i \omega \Delta_I \right)
\ee
with
\be
  \omega = \frac{\omega_1 + \omega_2}{2}\, ,\quad \omega_1 = \sqrt{q}\, ,\quad \omega_2 = \frac{1}{\sqrt{q}}\, ,
\ee
and
\be
  \Gamma_h (z) \equiv \Gamma_h (z;\, i \omega_1, i \omega_2)\, ,
\ee
where
\begin{align}
  \Gamma_h (z;\, \omega_1,\, \omega_2) & = \prod_{n_1,\, n_2 \geq 0} \frac{(n_1 + 1) \omega_1 + (n_2 + 1) \omega_2 - z}{n_1 \omega_1 + n_2 \omega_2 + z} \nonumber\\
  {} & = \textrm{exp} \left[i \int_0^\infty \frac{dx}{x} \left(\frac{z - \omega}{\omega_1 \omega_2 x} - \frac{\textrm{sin} (2 x (z - \omega))}{2\, \textrm{sin} (\omega_1 x)\, \textrm{sin} (\omega_2 x)} \right) \right]
\end{align}
defined for
\be
  0 < \textrm{Im} (z) < \textrm{Im} (\omega_1 + \omega_2)
\ee
is a hyperbolic gamma function \cite{MathThesis}. $\sigma_i$ parametrize the localization locus of the Coulomb branch. $k$ stands for the Chern-Simons level, and $I$ is the index for the chiral multiplets. $\alpha$ and $\rho$ denote the root of the adjoint representation and the weight of the representation $\mathcal{R}_I$ of the gauge group $G$ respectively. $\Delta_I$ is the R-charge of the scalar in the chiral multiplet. It turns out that the partition function $Z_{\textrm{singular space}} (q)$ equals the partition function of the same theory on a squashed three-sphere $\textrm{S}_b^3$ with $b = \sqrt{q}$.

\subsection{Results for ABJM Theory}

\begin{figure}[!htb]
      \begin{center}
        \includegraphics[width=0.5\textwidth]{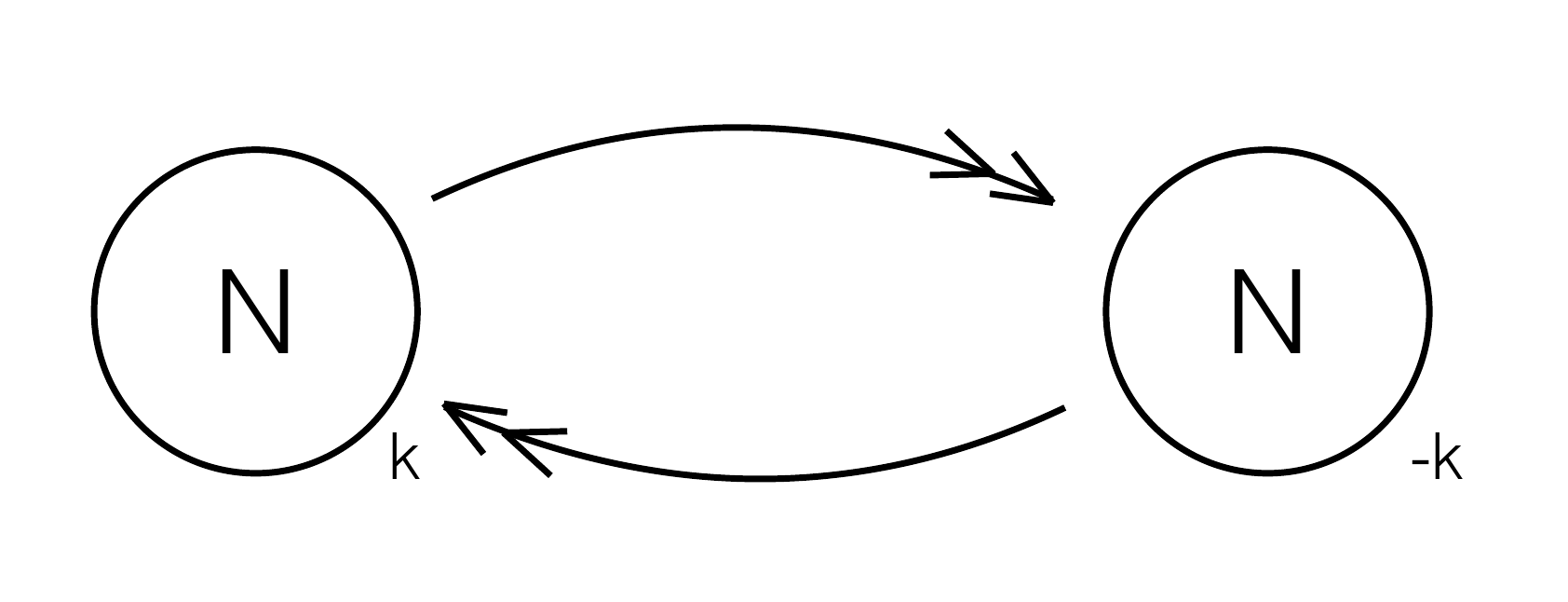}
      \caption{The quiver diagram for the ABJM theory}
      \label{fig:ABJM}
      \end{center}
    \end{figure}

As an example of the 3D superconformal field theory, the ABJM theory has been intensively studied. As first discussed by Aharony, Bergman, Jafferis and Maldacena in Ref.~\cite{ABJM}, the ABJM theory is a 3D $\mathcal{N}=6$ supersymmetric Chern-Simons-matter theory with the gauge group $U(N)_k \times U(N)_{-k}$, where $k$ stands for the Chern-Simons level. The theory describes the low-energy dynamics of $N$ M2-branes on $\mathbb{C}^4 / \mathbb{Z}_k$, and it has 4 bi-fundamental chiral multiplets, two of them in the $(N, \overline{N})$ representation and the other two in the $(\overline{N}, N)$ representation. The matter content of the ABJM theory can be illustrated using the quiver diagram in Fig.~\ref{fig:ABJM}.

With the development of the supersymmetric localization on curved manifolds \cite{Pestun}, the partition functions of some 3D supersymmetric gauge theories including the ABJM theory were studied in Ref.~\cite{Kapustin}, and they can be expressed as matrix integrals. In particular, the partition function of the ABJM theory is reduced to the following matrix model:
\be
  Z_{\textrm{ABJM}} = \int \prod_i d\sigma_i \, d\widetilde{\sigma}_i\, e^{- i k \pi (\sigma_i^2 - \widetilde{\sigma}_i^2)}\, \frac{\prod_{i < j} \left(4\, \textrm{sinh} (\pi \sigma_{ij}) \, \textrm{sinh} (\pi \widetilde{\sigma}_{ij}) \right)^2}{\prod_{i, j} \left( 2 \, \textrm{cosh} (\pi (\sigma_i - \widetilde{\sigma}_j))\right)^2}\, ,
\ee
where $\sigma_{ij} \equiv \sigma_i - \sigma_j$ and $\widetilde{\sigma}_{ij} \equiv \widetilde{\sigma}_i - \widetilde{\sigma}_j$ are the roots of $U(N)_k$ and $U(N)_{-k}$ respectively, and the weights in the representations $(N, \overline{N})$ and $(\overline{N}, N)$ are
\begin{align}
  \rho_{i, j}^{(N, \overline{N})} & = \sigma_i - \widetilde{\sigma}_j\, ,\nonumber\\
  \rho_{i, j}^{(\overline{N}, N)} & = - \sigma_i + \widetilde{\sigma}_j\, .
\end{align}

To evaluate these integrals, one still has to solve the matrix model, which sometimes can be nontrivial. To proceed the computation, on the one hand the matrix integral were evaluated directly under some approximations for the 3D superconformal field theories \cite{HerzogPufu, MartelliSparks}, while on the other hand using the Fermi gas approach one can obtain the final results of the perturbative contributions, which was done in Ref.~\cite{Marino}. The result for the partition function of the ABJM theory from the Fermi gas approach was obtained by Mari\~no and Putrov, which can be written in terms of the Airy function \cite{Marino}:
\be
  Z^{\textrm{ABJM}} \propto \textrm{Ai} \left[\left(\frac{\pi^2 k}{2} \right)^{1/3} \left(N - \frac{k}{24} - \frac{1}{3 k} \right) \right]\, .
\ee

As discussed above, the supersymmetric R\'enyi entropies of some 3D superconformal field theories can be expressed in terms of partition functions of these theories on squashed three sphere $\textrm{S}_b^3$, and these partition functions can still be written as matrix models using the technique of localization. For the ABJM theory, the partition function on a squashed three-sphere can be written as
\be
  Z^{\textrm{ABJM}}_{b^2} = \frac{1}{(N!)^2} \int \prod_i d\sigma_i \, d\widetilde{\sigma}_i\, e^{- i k \pi (\sigma_i^2 - \widetilde{\sigma}_i^2)}\, Z^{\textrm{vec}}_{b^2} \, Z^{\textrm{bi-fund}}_{b^2}\, ,
\ee
where
\begin{align}
  Z^{\textrm{vec}}_{b^2} & = \prod_{i < j} 4\, \textrm{sinh} (\pi b \sigma_{ij})\, \textrm{sinh} (\pi b^{-1} \sigma_{ij}) \, 4\, \textrm{sinh} (\pi b \widetilde{\sigma}_{ij})\, \textrm{sinh} (\pi b^{-1} \widetilde{\sigma}_{ij})\, , \nonumber\\
  Z^{\textrm{bi-fund}}_{b^2} & = \prod_{i, j} \frac{s_b \left(\sigma_i - \widetilde{\sigma}_j + i Q / 4 \right)^2}{s_b \left(\sigma_i - \widetilde{\sigma}_j - i Q / 4 \right)^2}\, ,
\end{align}
where $Q = b + 1/b$, and $s_b (x)$ is the double sine function. In the limit $b \to 1$,
\begin{align}
  Z^{\textrm{vec}}_{b^2=1} & = \prod_{i < j} \left(4\, \textrm{sinh} (\pi \sigma_{ij})\, \textrm{sinh} (\pi \widetilde{\sigma}_{ij}) \right)^2\, ,\nonumber\\
  Z^{\textrm{bi-fund}}_{b^2=1} & = \prod_{i, j} \frac{1}{\left(2\, \textrm{cosh} (\pi (\sigma_i - \widetilde{\sigma}_j)) \right)^2}\, ,
\end{align}
which reproduce the partition function of the ABJM theory on the round three-sphere found in Ref.~\cite{Kapustin}.

Recently, Hatsuda studied the partition function of the ABJM theory on a squashed three-spheres $\textrm{S}_b^3$ \cite{Hatsuda}, and found that for some cases the matrix model can be greatly simplified and evaluated analytically at large $N$ using the Fermi gas approach. For instance, when $k=1$ and $b^2 = 3$, the leading contribution to the partition function is
\be
  Z^{\textrm{ABJM}}_{b^2 = 3} = C_3^{-1/3}\, e^{A_3}\, \textrm{Ai} \left[C_3^{-1/3} (N - B_3) \right] + \cdots\, ,
\ee
where
\be
  A_3 = -\frac{\zeta(3)}{3 \pi^2} + \frac{\textrm{log}\, 3}{6}\, ,\quad B_3 = \frac{1}{8}\, ,\quad C_3 = \frac{9}{8 \pi^2}\, .
\ee
With these results, one can study the supersymmetric R\'enyi entropy at large $N$ beyond the leading order.

%%%%%%%%%%%%%%%%%%%%%%%%%%%%%%%%%%%%%%%%%%%
%%%%%%%%%%%%%%%%%%%%%%%%%%%%%%%%%%%%%%%%%%%
\section{Gravity Dual of Supersymmetric R\'enyi Entropy}\label{sec:GravDual}
%%%%%%%%%%%%%%%%%%%%%%%%%%%%%%%%%%%%%%%%%%%
%%%%%%%%%%%%%%%%%%%%%%%%%%%%%%%%%%%%%%%%%%%

The gravity dual of the supersymmetric R\'enyi entropies of 3D superconformal field theories (including the ABJM theory) has been constructed in Refs.~\cite{Rey, Nishioka-2}. Later, it was generalized to other dimensions \cite{4D5D-1, 4D5D-2, 5D6D-1, 5D6D-2, 2D3D-1, 2D3D-2, Yang-1, Yang-2}. In this section, we briefly review the gravity dual theory found in Refs.~\cite{Rey, Nishioka-2}.

As discussed in Ref.~\cite{Nishioka-1}, due to the conical singularity one has to turn on a R-symmetry gauge field in order to preserve supersymmetry. In the spirit of the AdS/CFT correspondence, instead of finding an AdS space with the branched sphere as the boundary, one can first perform the conformal transformation introduced in Section~\ref{sec:introduction} to the branched sphere, which maps the branched three-sphere into $\textrm{S}^1 \times \mathbb{H}^2$, i.e.
\begin{align}
  ds^2 & = d\theta^2 + q^2 \, \textrm{sin}^2 \theta\, d\tau^2 + \textrm{cos}^2 \theta\, d\phi^2 \nonumber\\
  {} & = \textrm{sin}^2 \theta \left[d\tilde{\tau}^2 + du^2 + \textrm{sinh}^2 u \, d\phi^2 \right]\, ,
\end{align}
where $\tau \in [0,\, 2 \pi)$, $\tilde{\tau} = q \tau \in [0,\, 2 q \pi)$, and
\be\label{eq:ConfTrafo}
  \textrm{sinh}\, u = - \textrm{cot}\, \theta\, .
\ee
Next, one can find an $\textrm{AdS}_4$ topological black hole with the metric \cite{Brill}:
\be\label{eq:AdS4metric}
  ds^2 = - f(r)\, dt^2 + \frac{1}{f(r)} dr^2 + r^2\, d\Sigma (\mathbb{H}^2)\, ,
\ee
whose boundary is $\mathbb{R}^1 \times \mathbb{H}^2$, where
\be
  f(r) = \frac{r^2}{L^2} - 1 - \frac{2m}{r} + \frac{Q^2}{r^2}\, ,
\ee
and
\be
  d\Sigma (\mathbb{H}^2) = du^2 + \textrm{sinh}^2 u\, d\phi^2\, .
\ee
This metric can be viewed as solutions to the 4D $\mathcal{N}=2$ gauged supergravity given by the effective action \cite{FreedmanDas}, whose bosonic part is
\be
  I = -\frac{1}{2 \, \ell_P^2} \int d^4 x\, \sqrt{-g} \left(2 \Lambda + R - \frac{1}{g^2} F_{\mu\nu} F^{\mu\nu} \right)\, .
\ee
The gauge field is given by
\be
  A = \left(\frac{Q}{r} - \frac{Q}{r_h} \right) dt\, ,
\ee
where $r_h$ is the horizon radius of the black hole determined by $f(r_h) = 0$.

As explained in Refs.~\cite{Ortin, Rey}, to preserve the supersymmetry, the condition
\be
  m^2 + Q^2 = 0
\ee
holds for both the charged case ($Q \neq 0$) and the neutral case ($Q = 0$).

For $Q^2 = - m^2 \neq 0$,
\be
  f(r) = \frac{r^2}{L^2} - \left(1 + \frac{m}{r} \right)^2\, ,
\ee
the metric \eqref{eq:AdS4metric} corresponds to a charged topological black hole. As shown in Refs.~\cite{Rey, Nishioka-2}, the Bekenstein-Hawking entropy of the charged topological black hole equals the R\'enyi entropy of the superconformal field theory on the boundary. In particular, the result from the gravity dual recovers the relation between the R\'enyi entropy $S_q$ and the entanglement entropy $S_1$ for the 3D superconformal field theories:
\be
  S_q = \frac{3 q + 1}{4 q} S_1\, .
\ee
For $Q = m = 0$,
\be
  f(r) = \frac{r^2}{L^2} - 1\, ,
\ee
and $r_h = L$. The gravity solution is dual to a 3D superconformal field theory on $\textrm{S}^1 \times \mathbb{H}^2$ with $q=1$. Correspondingly, the black hole entropy in this case equals the entanglement entropy of the boundary superconformal field theory. The evaluation of the gravity free energy at classical level supports this identification.

As we know, the Bekenstein-Hawking entropy corresponds to the classical result of the gravity. Using supersymmetric localization, we can obtain more precise result and go beyond the classical result. Hence, in this way we can test the gravity dual and the AdS/CFT correspondence more precisely.

In this paper, we consider the neutral topological black hole, whose entropy gives the entanglement entropy of the superconformal field theory on the boundary. For this case, the branching parameter $q=1$, which corresponds to the round three sphere. One can nevertheless perform the conformal transformation \eqref{eq:ConfTrafo}. The hyperbolic $\textrm{AdS}_4$ space becomes an $\textrm{AdS}_4$ neutral topological black hole. The entanglement entropy of the superconformal field theory on the boundary is supposed to be equal to the bulk black hole entropy. The equality can be tested more precisely using the results of the localization of supergravity.

%%%%%%%%%%%%%%%%%%%%%%%%%%%%%%%%%%%%%%%%%%%
%%%%%%%%%%%%%%%%%%%%%%%%%%%%%%%%%%%%%%%%%%%
\section{4D $\mathcal{N}=2$ Off-Shell Gauged Supergravity and Its Localization}\label{sec:SUGRAloc}
%%%%%%%%%%%%%%%%%%%%%%%%%%%%%%%%%%%%%%%%%%%
%%%%%%%%%%%%%%%%%%%%%%%%%%%%%%%%%%%%%%%%%%%

In this section, we discuss the localization of the 4D $\mathcal{N}=2$ off-shell supergravity on $\textrm{AdS}_4$ topological black hole with the boundary $\textrm{S}^1 \times \mathbb{H}^2$. The steps are similar to the ones in Ref.~\cite{4DSUGRAloc}, however, there are some subtle differences which consequently lead to different final results.

\subsection{4D $\mathcal{N}=2$ Off-Shell Gauged Supergravity}

The 4D $\mathcal{N}=2$ off-shell supergravity theory can be obtained as a consistent truncation of M-theory on a Sasaki-Einstein manifold $X_7$. The theory was originally constructed in Ref.~\cite{deWit} and also reviwed in Ref.~\cite{4DSUGRAloc}. We also briefly summarize the theory in Appendix~\ref{app:SUGRA}.

The $\mathcal{N}=2$ superconformal algebra has the generators:
\be
  P_a,\, M_{ab},\, D,\, K_a,\, Q_i,\, S^i,\, U_{ij},
\ee
which correspond to the generators of translations, Lorentz rotations, dilatations, special conformal transformations, usual supersymmetry transformations, special conformal supersymmetry transformations and the 4D $SU(2)$ R-symmetry respectively. The gauge fields corresponding to these generators are
\be
  e_\mu^a,\, \omega_\mu^{ab},\, b_\mu,\, f_\mu^a,\, \psi_\mu^i,\, \phi_\mu^i,\, \mathcal{V}_\mu^{ij}
\ee
respectively.

To construct a 4D off-shell $\mathcal{N}=2$ supergravity, one needs the Weyl multiplet $\bold{W}$:
\be
  \bold{W} = \big( e_\mu^a,\, \psi_\mu^i,\, b_\mu,\, A_\mu,\, \mathcal{V}_{\mu j}^i,\, T_{ab}^{ij},\, \chi^i,\, D \big)\, ,
\ee
the vector multiplet $\bold{X}^I$:
\be
  \bold{X}^I = \big( X^I,\, \Omega_i^I,\, W_\mu^I,\, Y_{ij}^I \big)\, ,
\ee
and the hypermultiplet $(A_i\,^\alpha,\, \zeta^\alpha)$. More details about these multiplets and their supersymmetric transformations can be found in Appendix~\ref{app:SUGRA}.

Given a prepotential $F(X)$, the two-derivative off-shell action for the bosonic fields is given by
\begin{align}
  S & = \int d^4 x \sqrt{g} \Bigg[ N_{IJ} \overline{X}^I X^J \left(\frac{R}{6} + D \right) + N_{IJ} \partial \overline{X}^I \partial X^J - \frac{1}{8} N_{IJ} Y^{ij I} Y_{ij}^J \nonumber\\
  {} & + \Bigg(- \nabla A^i\,_\beta \nabla A_i\,^\alpha - \left(\frac{R}{6} - \frac{D}{2} \right) A^i\,_\beta A_i\,^\alpha + F^i\,_\beta F_i\,^\alpha + 4 g^2 A^i\,_\beta \overline{X}^\alpha\,_\gamma X^\gamma\,_\delta A_i\,^\delta \nonumber\\
  {} & \qquad + g A^i\,_\beta (Y^{jk})_\gamma\,^\alpha A_k\,^\gamma \epsilon_{ij} \Bigg) d_\alpha\,^\beta \Bigg]\, ,\label{eq:SUGRAaction}
\end{align}
where
\be
  N_{IJ} \equiv \frac{1}{2 i} \left(F_{IJ} - \overline{F}_{IJ} \right)\, ,\quad F_{IJ} \equiv \partial_I \partial_J F(X)\, ,
\ee
and $F_i\,^\alpha$ is related to the field $A_i\,^{\alpha (z)}$ discussed in Appendix~\ref{app:SUGRA} in the following way:
\be
  F_i\,^\alpha = a A_i\,^{\alpha (z)}\, .
\ee
The term $\sim R A^2$ provides a negative cosmological constant for the $\textrm{AdS}_4$ space.

\subsection{Localization of Supergravity}

As discussed before, to find the gravity dual of the supersymmetric R\'enyi entropy, one can perform a conformal transformation \eqref{eq:ConfTrafo} on the boundary, which maps the branched three-sphere into $\textrm{S}^1 \times \mathbb{H}^2$. Correspondingly, the metric in the bulk now should be the $\textrm{AdS}_4$ topological black hole given by the metric \eqref{eq:AdS4metric} to match the boundary.

In this section, we discuss the localization of the 4D $\mathcal{N}=2$ off-shell gauged supergravity on the background of the $\textrm{AdS}_4$ neutral topological black hole \eqref{eq:AdS4metric}. In other words, we focus on the case with the branching parameter $q=1$, which corresponds to the entanglement entropy of the ABJM theory across a circle $\textrm{S}^1$ on the boundary. The discussions are similar to the supergravity localization on the hyperbolic $\textrm{AdS}_4$ in Ref.~\cite{4DSUGRAloc}.

\subsubsection{BPS Equations}\label{sec:BPSEq}

Let us first consider the BPS equations for various supergravity multiplets. For the Weyl multiplet, by setting $\delta \psi_\mu\,^i = 0$ one obtains the Killing spinor equation:
\be\label{eq:KillingEq}
  2 \nabla_\mu \epsilon^i + i A_\mu \epsilon^i - \frac{1}{8} T_{ab}\,^{ij} \gamma^{ab} \gamma_\mu \epsilon_j = \gamma_\mu \eta^i\, ,
\ee
where we set the background $b_\mu = \mathcal{V}_\mu\,^{ij} = 0$, and
\be
  \nabla_\mu \epsilon^i \equiv \partial_\mu \epsilon^i + \frac{1}{4} \omega_\mu\,^{ab} \gamma_{ab}\, \epsilon^i\, .
\ee
For $q=1$, one can further set $A_\mu = T_{ab}\,^{ij} = 0$ in Eq.~\eqref{eq:KillingEq}, and the Killing spinor equation becomes
\be\label{eq:KillingEqSimplified}
  \nabla_\mu \epsilon^i = \frac{1}{2} \gamma_\mu \eta^i\, .
\ee

The upper and the lower indices for $\epsilon$ denote the positive and the negative chirality respectively, and the opposite for $\eta$. We can use the Dirac notation to combine different components into Dirac spinors
\be
  \xi = (\xi_+^i,\, \xi_-^i)\, ,\quad \eta = (\eta_+^i,\, \eta_-^i)\, ,
\ee
where
\be
  \xi_+^i \equiv \epsilon^i\, ,\quad \epsilon_i \equiv i \epsilon_{ij} \xi_-^j\, ,\quad \eta_i \equiv - \epsilon_{ij} \eta_+^j\, ,\quad \eta^i \equiv i \eta_-^i\, .
\ee
Using these notations, we can rewrite the Killing spinor equation \eqref{eq:KillingEqSimplified} for $q=1$ as follows:
\be\label{eq:KillingEqTemp}
  \nabla_\mu \xi^i = \frac{i}{2} \gamma_\mu \eta^i\, .
\ee

We would like to recover the Killing spinor equation for the topological black hole discussed in Ref.~\cite{Rey} for $q=1$.
To do so, let us first consider the general Killing spinor equation for the $\textrm{AdS}_4$ topological black hole \cite{Rey}:
\be
  \nabla_\mu \epsilon - i g A_\mu \epsilon + \frac{i}{4} F_{ab} \gamma^{ab} \gamma_\mu \epsilon = -\frac{1}{2} g \gamma_\mu \epsilon\, ,
\ee
where
\be
  A = \left(\frac{Q}{r} - \frac{Q}{r_h} \right) dt
\ee
with $r_h$ denoting the position of the horizon, and consequently only the components $F_{rt}$ and $F_{tr}$ are nonvanishing. For the branching paramter $q=1$ considered in this paper, the black hole is neutral, i.e. $Q=0$, hence both $A_\mu$ and $F_{\mu\nu}$ vanish for this case. As discussed in Appendix~\ref{app:Killing}, using the charge conjugation matrix $B$ one can define the charge conjugate spinor satisfying another Killing spinor equation:
\be
  \nabla_\mu \epsilon^c + i g A_\mu \epsilon^c + \frac{i}{4} F_{ab} \gamma^{ab} \gamma_\mu \epsilon^c = \frac{1}{2} g \gamma_\mu \epsilon^c\, .
\ee
The coupling constant $g$ is related to the AdS radius \cite{Rey}:
\be
  L = \frac{1}{g}\, .
\ee
For $q=1$, the two Killing spinor equations are
\be\label{eq:ReyKillingSp}
  \nabla_\mu \epsilon = -\frac{1}{2 L} \gamma_\mu \epsilon\, ,\quad \nabla_\mu \epsilon^c = \frac{1}{2 L} \gamma_\mu \epsilon^c\, ,
\ee
which can be written into a more compact form using the Dirac notation:
\be
  \nabla_\mu \widetilde{\xi}^i = \frac{1}{2L} \gamma_\mu (\sigma_3)^i\,_j \widetilde{\xi}^j\, ,
\ee
with
\be
  \widetilde{\xi}^1 \equiv \epsilon^c\, ,\quad \widetilde{\xi}^2 \equiv \epsilon\, .
\ee
Defining
\be
  \xi^i \equiv \frac{1 + i \gamma_5}{2}\, \widetilde{\xi}^i\, ,
\ee
one can further obtain an equivalent expression for the Killing spinor equation at $q=1$:
\be\label{eq:KillingEqFinal}
  \nabla_\mu \xi^i = \frac{i}{2 L} \gamma_5 \gamma_\mu (\sigma_3)^i\,_j \xi^j\, .
\ee
Comparing this equation with Eq.~\eqref{eq:KillingEqTemp}, we should identify
\be\label{eq:etaxi}
  \eta^i = - \frac{1}{L} \gamma_5 (\sigma_3)^i\,_j \xi^j\, .
\ee
Eq.~\eqref{eq:KillingEqFinal} will be the Killing spinor equation used throughout the rest of this paper.

Next, for the vector multiplet, the BPS equations are obtained from $\delta \Omega_i = 0$. Setting $F_{\mu\nu} = T_{\mu\nu} = 0$ and distinguishing different chiralities, we obtain
\begin{align}
  \delta \Omega_+^i & = -i \slashed \partial X \xi_-^i - \frac{1}{2} Y^i\,_j \xi_+^j + X \eta_+^i = 0\, ,\nonumber\\
  \delta \Omega_-^i & = -i \slashed \partial \overline{X} \xi_+^i - \frac{1}{2} Y^i\,_j \xi_-^j + \overline{X} \eta_-^i = 0\, ,
\end{align}
which can be combined into
\be
  -i \slashed \partial (H - i \gamma_5 J) \xi^i - \frac{1}{2} Y^i\,_j \xi^j - \frac{1}{L} (H + i \gamma_5 J) \gamma_5 (\sigma_3)^i\,_j \xi^j = 0\, ,
\ee
where we have parametrized $X = H + i J$ and used the expression for $\eta^i$ given above. For constant $H$ and $J$, the BPS equations above have the solution:
\be
  H = 0\, ,\quad Y^1\,_1 = - Y^2\,_2 = - \frac{2i}{L} J\, ,\quad Y^1\,_2 = - Y^2\,_1 = 0\, .
\ee

The BPS equation for the hypermultiplet can be obtained by setting the modified supersymmetric transformation $\delta \zeta = 0$ (see Appendix~\ref{app:SUGRA}), which leads to
\begin{align}
  \delta \zeta_{\alpha +} & = i \slashed \nabla A^i\,_\alpha \epsilon_{ij} \xi_-^j + 2 g \overline{X}_\alpha\,^\beta A^i\,_\beta \epsilon_{ij} \xi_+^j - A^i\,_\alpha \epsilon_{ij} \eta_+^j + F^i\,_\alpha \epsilon_{ij} \xi_+^j = 0\, , \nonumber\\
  \delta \zeta_-^\alpha & = \slashed \nabla A_i\,^\alpha \xi_+^i - 2 g i X^\alpha\,_\beta A_i\,^\beta \xi_-^i + i A_i\,^\alpha \eta_-^i - i F_i\,^\alpha \xi_-^i = 0\, ,
\end{align}
where $F_i\,^\alpha \equiv a A_i\,^{\alpha (z)}$ satisfying $F_i\,^\alpha = (F^i\,_\alpha)^* = \epsilon_{ij} \epsilon^{\alpha\beta} F^j\,_\beta$. One can combine these two equations using the Dirac notation in the following way:
\be
  \slashed \nabla A_i\,^\alpha \xi^i - 2 g i (H^I - i \gamma_5 J^I) (t_I)^\alpha\,_\beta A_i\,^\beta \xi^i + i A_i\,^\alpha \eta^i - i F_i\,^\alpha \xi^i = 0\, .
\ee
We consider the model with the charges
\be
  t_I A_i\,^\alpha = P_I (i \sigma_3)^\alpha\,_\beta A_i\,^\beta\, ,
\ee
where $P_I$ are moment maps on the hyperk\"ahler manifold with the scalars in the hypermultiplet as sections. In the gauge $A^i\,_\alpha \propto \delta^i\,_\alpha$, using the relation \eqref{eq:etaxi} one can express the BPS equation for the hypermultiplet as
\be
  \left[2g (H\cdot P) - 2 g i \gamma_5 (J\cdot P) - \frac{i}{L} \gamma_5 \right]  A_i\,^\alpha (\sigma_3)^i\,_j \xi^j - i F_i\,^\alpha \xi^i = 0\, ,
\ee
which leads to the solution
\be\label{eq:BPSsolhyper}
  F_j\,^\alpha = - 2 i g A_i\,^\alpha (\sigma_3)^i\,_j (H\cdot P)\, ,\quad 2 g (J\cdot P) = - \frac{1}{L}\, .
\ee

\subsubsection{Attractor Solution}
As we discussed before, given a prepotential $F(X)$, the two-derivative off-shell action for the bosonic fields is given by Eq.~\eqref{eq:SUGRAaction}. Now let us take a closer look at the theory and analyze its attractor solution. Later in the localization procedure, the localization locus will fluctuate around the attractor solution discussed in this subsection.

First, the field $D$ plays the role of a Lagrange multiplier, which imposes the condition:
\be\label{eq:attTemp-1}
  N_{IJ} \overline{X}^I X^J + \frac{1}{2} A^i\,_\beta A_i\,^\alpha d_\alpha\,^\beta = 0\, .
\ee
By requiring that the terms containing the Ricci scalar reproduce the Einstein-Hilbert action, we obtain
\be\label{eq:attTemp-2}
  \frac{1}{6} N_{IJ} \overline{X}^I X^J - \frac{1}{6} A^i\,_\beta A_i\,^\alpha d_\alpha\,^\beta = \frac{1}{16 \pi G}\, ,
\ee
where $G$ is the Newton's constant. The equations \eqref{eq:attTemp-1} \eqref{eq:attTemp-2} lead to
\be\label{eq:attsol}
  N_{IJ} \overline{X}^I X^J = \frac{1}{8 \pi G}\, ,\quad A^i\,_\beta A_i\,^\alpha d_\alpha\,^\beta = -\frac{1}{4 \pi G}\, .
\ee
In the gauge $A_i\,^\alpha \propto \delta_i^\alpha$, the second equation above implies
\be\label{eq:attsolA}
  A_i\,^\alpha = \frac{1}{\sqrt{8 \pi G}} \delta_i^\alpha\, ,
\ee
where we have used $d_\alpha\,^\beta = - \delta_\alpha\,^\beta$ discussed in Appendix~\ref{app:SUGRA}.

By analyzing the field equations of various fields in the action \eqref{eq:SUGRAaction}, we arrive at the same solution that we found before from the BPS equations \eqref{eq:BPSsolhyper}:
\be
  2 g (J \cdot P) = - \frac{1}{L}\, ,
\ee
more precisely,
\be\label{eq:attSol-1}
  8 g J^0 P_0 = - \frac{1}{L}\, ,\quad 8 g J^1 P_1 = - \frac{3}{L}\, .
\ee

For the prepotential $F(X) = \sqrt{X^0 (X^1)^3}$, the first one of Eq.~\eqref{eq:attsol} becomes
\be
  \frac{1}{4 i} |X^0|^2 \left(\sqrt{\frac{X^1}{X^0}} - \overline{\sqrt{\frac{X^1}{X^0}}} \right)^3 = \frac{1}{8 \pi G}\, ,
\ee
which consequently leads to
\be\label{eq:attSol-2}
  (J^0)^{1/2} (J^1)^{3/2} = \frac{i}{16 \pi G}\, ,
\ee
where we choose $\sqrt{-1} = -i$.

\subsubsection{Localization Action}\label{sec:LocAction}

As in the standard localization procedure, we can add a SUSY-exact term to the action without changing the partition function of the theory. The SUSY-exact term is called the localization action. In our case, we choose the following localization action for the vector multiplet:
\be
  \Delta S = \delta ((\delta \Omega)^\dagger \Omega)\, ,
\ee
where $\Omega$ denotes the gaugino field in the vector multiplet. The bosonic part of the localization action is
\be\label{eq:LocAction}
  (\Delta S)_{\textrm{bos}} = (\delta \Omega)^\dagger \delta \Omega\, .
\ee
We can solve $(\Delta S)_{\textrm{bos}} = 0$ to find the localization locus. Some details are presented in Appendix~\ref{app:LocAction}.

When expanding the localization action, we choose the Killing spinor found in Ref.~\cite{Rey} for the topological black hole with $q=1$:
\be
  \epsilon = e^{-\frac{i}{2 q L} \tau_E}\, e^{i \frac{u}{2} \gamma_4 \gamma_1 \gamma_2}\, e^{\frac{\phi}{2} \gamma_{23}}\, \tilde{\epsilon}(r)
\ee
with
\be
  \tilde{\epsilon}(r) = \left(\sqrt{\frac{r}{L} + \sqrt{f(r)}} - i \gamma_4 \sqrt{\frac{r}{L} - \sqrt{f(r)}} \right) \left(\frac{1 - \gamma_1}{2} \right) \epsilon'_0\, ,
\ee
where $\epsilon'_0$ is an arbitrary constant spinor, and $f(r)$ is the factor appearing in the metric of the topological black hole \eqref{eq:AdS4metric}:
\begin{displaymath}
  ds^2 = - f(r)\, dt^2 + \frac{1}{f(r)} dr^2 + r^2\, d\Sigma (\mathbb{H}^2)\, .
\end{displaymath}
In principle, there are 8 independent Killing spinors $\epsilon$. Moreover, $\epsilon$ and $\epsilon^c \equiv B \epsilon^*$ satisfy the Killing spinor equations \eqref{eq:ReyKillingSp}:
\begin{displaymath}
  \nabla_\mu \epsilon = -\frac{1}{2 L} \gamma_\mu \epsilon\, ,\quad \nabla_\mu \epsilon^c = \frac{1}{2 L} \gamma_\mu \epsilon^c\, .
\end{displaymath}
As we discussed in Subsection~\ref{sec:BPSEq}, it is more convenient to work with the Killing spinors
\be
  \xi^i \equiv \frac{1 + i \gamma_5}{2}\, \widetilde{\xi}^i\, ,
\ee
with
\be
  \widetilde{\xi}^1 \equiv \epsilon^c\, ,\quad \widetilde{\xi}^2 \equiv \epsilon\, .
\ee
and they satisfy the equivalent Killing spinor equation \eqref{eq:KillingEqFinal}:
\begin{displaymath}
  \nabla_\mu \xi^i = \frac{i}{2 L} \gamma_5 \gamma_\mu (\sigma_3)^i\,_j \xi^j\, .
\end{displaymath}
The Killing spinors $\xi^i$ generate the Killing vector
\be
  v = \xi^\dagger \gamma^\mu \xi \partial_\mu = \mathcal{L}_{U(1)}\, ,
\ee
which is a linear combination of the compact $U(1)$'s along the compact directions $\tau$ and $\phi$ in the metric \eqref{eq:EuclideanAdS4metric}.

Using the Killing spinor discussed above with the special choice of the constant spinor $\epsilon'_0 = (1, 0, i, 0)^T$, we can compute various Killing spinor bilinears, and expand the localization action \eqref{eq:LocAction} explicitly. The localization action \eqref{eq:LocAction} can be expressed as a sum of some squares \eqref{eq:SquareSum}. By requiring these squares vanish, we obtain the following solutions:
\be
  H = \frac{C}{\textrm{cosh} (\eta)}\, ,\quad Y^1\,_1 = \frac{2 C}{\textrm{cosh}^2 (\eta)}\, ,\quad \textrm{ for } u = 0\, ;
\ee
\be
  J = \textrm{const}\, ,\quad F_{ab} V^b = 0\, ,
\ee
where $C$ is an arbitrary constant, and the constant value of $J$ is fixed by the attractor solutions \eqref{eq:attSol-1} \eqref{eq:attSol-2}. Together with the BPS solutions found in Subsection~\ref{sec:BPSEq}, these form the localization locus:
\begin{align}
  X^I & = H^I + i J^I = \frac{C^I}{\textrm{cosh} (\eta)} + i J^I = \frac{J^I h^I}{\textrm{cosh} (\eta)} + i J^I\, , \nonumber\\
  (Y^I)^1\,_1 & = - (Y^I)^2\,_2 = \frac{2 C^I}{\textrm{cosh}^2 (\eta)} - \frac{2 i}{L} J^I = \frac{2 J^I h^I}{\textrm{cosh}^2 (\eta)} - \frac{2 i}{L} J^I\, ,\label{eq:LocLocus}
\end{align}
where we have written the gauge index $I$ explicitly and used the parametrization $C^I = J^I h^I$, and again the values of $J^I$ are fixed to be the attractor solutions given by Eq.~\eqref{eq:attSol-1} and Eq.~\eqref{eq:attSol-2}.

For the hypermultiplet, as discussed in Appendix~\ref{app:LocAction}, we require $\delta \zeta = 0$ for all 8 Killing spinors, which leads to the solutions
\be
  F_i\,^\alpha = - \frac{2 i g}{\sqrt{8 \pi G}} (\sigma_3)^\alpha\,_j (H\cdot P)\, ,\quad 2 g (J\cdot P) = - \frac{1}{L}
\ee
with $F_i\,^\alpha$ and $P_I$ given by
\be
  F_i\,^\alpha = a A_i\,^{\alpha (z)}\, ,\quad t_I A_i\,^\alpha = P_I (i \sigma_3)^\alpha\,_\beta A_i\,^\beta\, .
\ee
These solutions coincide with the solutions \eqref{eq:BPSsolhyper} to the BPS equations under the attractor solution \eqref{eq:attsolA}.

\subsubsection{Action on Localization Locus}\label{sec:Action}

Now we would like to evaluate the action \eqref{eq:SUGRAaction} at the localization locus \eqref{eq:LocLocus} obtained in the previous subsection. We distinguish the action for the vector multiplet and the action for the hypermultiplet:
\begin{align}
  S_{\textrm{vec}} & = \int d^4 x \sqrt{g} \Bigg[ N_{IJ} \overline{X}^I X^J \frac{R}{6} + N_{IJ} \partial \overline{X}^I \partial X^J - \frac{1}{8} N_{IJ} Y^{ij I} Y_{ij}^J \Bigg]\, ,\\
  S_{\textrm{hyp}} & = \int d^4 x \sqrt{g} \Bigg[\Bigg( - \frac{R}{6} A^i\,_\beta A_i\,^\alpha + F^i\,_\beta F_i\,^\alpha + 4 g^2 A^i\,_\beta \overline{X}^\alpha\,_\gamma X^\gamma\,_\delta A_i\,^\delta \nonumber\\
  {} & \qquad\qquad\qquad + g A^i\,_\beta (Y^{jk})_\gamma\,^\alpha A_k\,^\gamma \epsilon_{ij} \Bigg) d_\alpha\,^\beta \Bigg]\, .
\end{align}
However, there is a subtle difference between the $\textrm{AdS}_4$ case considered in Ref.~\cite{4DSUGRAloc} and the case considered in this paper. In Ref.~\cite{4DSUGRAloc}, the $\textrm{AdS}_4$ metric is given by
\be
  ds^2 = L^2 (d\eta^2 + \textrm{sinh}^2 (\eta)\, d\Omega_3^2)\, ,
\ee
which leads to $d^4 x\, \sqrt{g} = L^4 d\Omega_3 dr (r^2 - 1)$ with $r \equiv \textrm{cosh} (\eta)$. In our case, the Euclidean $\textrm{AdS}_4$ neutral topological black hole is given by the metric (see Appendix~\ref{app:Killing}):
\be
  ds^2 = f(r)\, d\tau^2 + \frac{1}{f(r)}\, dr^2 + r^2 \, d\Sigma(\mathbb{H}^2)\, ,
\ee
where
\be
  f(r) = \frac{r^2}{L^2} - 1\, ,
\ee
therefore, the measure becomes $d^4 x \sqrt{g} = d\tau\, d\Sigma_2\, dr\, r^2$. Besides the volume form, we see that the measure for the integral $dr$ differs for the two cases.

Although in our case the integrand evaluated at the localization locus is the same as the $\textrm{AdS}_4$ case discussed in Ref.~\cite{4DSUGRAloc}, the final result is not the same due to the difference in the measure. Some details of the computation are presented in Appendix~\ref{app:Action}. The final results for $S_{\textrm{vec}}$ and $S_{\textrm{hyp}}$ are
\begin{align}
  S_{\textrm{vec}} & = \frac{\Omega_3^{\textrm{reg}} L^2}{32 \pi G} \Bigg[- 4 r_0^3 + \frac{r_0}{2} \left((h^0)^2 - 3 h^1 (4 i + h^1) - 2 h^0 (2 i + 3 h^1) \right) \nonumber\\
  {} & \qquad\qquad\quad - 2 (h^1 - i)^{3/2} (h^0 - i)^{1/2} + 6 (h^1 + i)^{3/2} (h^0 + i)^{1/2} \nonumber\\
  {} & \qquad\qquad\quad - 6 i (h^1 + i)^{1/2} (h^0 + i)^{1/2} - 2 i (h^1 + i)^{3/2} (h^0 + i)^{-1/2} + \mathcal{O} (1 / r_0) \Bigg]\, ,\\
  S_{\textrm{hyp}} & = \frac{i \Omega_3^{\textrm{reg}}\, L^2}{16 \pi G} (r_0 - 1) \left(h^0 + 3 h^1 \right)\, ,
\end{align}
where $r_0$ and $\Omega_3^{\textrm{reg}}$ are the cutoff and the regularized volume of the boundary $\textrm{S}^1 \times \mathbb{H}^2$ respectively, and we have used the attractor solutions given by Eq.~\eqref{eq:attSol-1}:
\begin{displaymath}
  8 g J^0 P_0 = - \frac{1}{L}\, ,\quad 8 g J^1 P_1 = - \frac{3}{L}\, ,
\end{displaymath}
and Eq.~\eqref{eq:attSol-2}:
\begin{displaymath}
  (J^0)^{1/2} (J^1)^{3/2} = \frac{i}{16 \pi G}\, .
\end{displaymath}
Altogether, the action evaluated at the localization locus is
\begin{align}
  S & = S_{\textrm{vec}} + S_{\textrm{hyp}} \nonumber\\
  {} & = - \frac{\Omega_3^{\textrm{reg}} L^2}{32 \pi G} \Bigg[4 r_0^3 - \frac{r_0}{2} \left((h^0)^2 - 3 (h^1)^2 - 6 h^0 h^1 \right) \nonumber\\
  {} & \qquad\qquad\quad + 2 (h^1 - i)^{3/2} (h^0 - i)^{1/2} - 6 (h^1 + i)^{3/2} (h^0 + i)^{1/2} + 2 i (h^0 + 3 h^1) \nonumber\\
  {} & \qquad\qquad\quad + 6 i (h^1 + i)^{1/2} (h^0 + i)^{1/2} + 2 i (h^1 + i)^{3/2} (h^0 + i)^{-1/2} + \mathcal{O} (1 / r_0) \Bigg]\, .\label{eq:ActionOnLocus}
\end{align}

\subsubsection{Holographic Renormalization}

To remove the divergence depending on the cutoff $r_0$ in the action \eqref{eq:ActionOnLocus}, which is
\be\label{eq:divergence}
  - \frac{\Omega_3^{\textrm{reg}} L^2}{8 \pi G} \Bigg[r_0^3 - \frac{r_0}{8} \left((h^0)^2 - 3 (h^1)^2 - 6 h^0 h^1 \right) \Bigg]\, ,
\ee
we apply the standard holographic renormalization by adding some boundary counter-terms:
\be
  S_{\textrm{ct}} = S_{\textrm{GH}} + \frac{1}{2} S_{B2}\, .
\ee
$S_{\textrm{GH}}$ is the Gibbons-Hawking term given by
\begin{align}
  S_{\textrm{GH}} & = \int d^3 x\, \sqrt{g_3}\, N_{IJ} \overline{X}^I X^J \frac{\kappa}{3} \nonumber\\
  {} & = \Omega_3^{\textrm{reg}} L^2 \left(r_0^3 - \frac{2}{3} r_0 \right) \left(N_{IJ} \overline{X}^I X^J\right) \bigg|_{r_0} \nonumber\\
  {} & = \frac{\Omega_3^{\textrm{reg}} L^2}{8 \pi G} \left(r_0^3 - \frac{2}{3} r_0 \right) \left[1 - \frac{(h^0)^2 - 3 (h^1)^2 - 6 h^0 h^1}{8 r_0^2} + \mathcal{O} (r_0^{-4})\right]\, ,
\end{align}
where $\kappa$ is the extrinsic curvature, which for the metric \eqref{eq:EuclideanAdS4metric} has the value:
\be
  \kappa = \frac{f'(r)}{2 \sqrt{f(r)}} + \frac{2 \sqrt{f(r)}}{r}\, .
\ee
$S_{B2}$ is the boundary term proportional to the boundary scalar curvature: \footnote{Compared to Ref.~\cite{4DSUGRAloc}, in this paper there is an extra factor $4$ in $S_{B2}$, because for the $\textrm{AdS}_4$ case considered in Ref.~\cite{4DSUGRAloc} the boundary scalar curvature is $24 / (L^2 \, \textrm{sinh}^2 (\eta))$, which is $6 / (L^2 \, \textrm{sinh}^2 (\eta))$ according to our convention.}
\begin{align}
  S_{B2} & = - \int d^3 x\, \sqrt{g_3} N_{IJ} \overline{X}^I X^J \frac{2 L \mathcal{R}}{3} \nonumber\\
  {} & = \Omega_3^{\textrm{reg}} L^2 \left(\frac{4}{3} r_0 + \mathcal{O} (r_0^{-1}) \right) \left(N_{IJ} \overline{X}^I X^J\right) \bigg|_{r_0} \nonumber\\
  {} & = \frac{\Omega_3^{\textrm{reg}} L^2}{8 \pi G} \left(\frac{4}{3} r_0 + \mathcal{O} (r_0^{-1}) \right) \left[1 - \frac{(h^0)^2 - 3 (h^1)^2 - 6 h^0 h^1}{8 r_0^2} + \mathcal{O} (r_0^{-4})\right]\, ,
\end{align}
where the boundary scalar curvature is $\mathcal{R} = -2 / r^2$ for the boundary $\textrm{S}^1 \times \mathbb{H}^2$. Therefore,
\begin{align}
  S_{\textrm{ct}} & = S_{\textrm{GH}} + \frac{1}{2} S_{B2} \nonumber\\
  {} & =  \frac{\Omega_3^{\textrm{reg}} L^2}{8 \pi G} \left[r_0^3 - \frac{r_0}{8} \left((h^0)^2 - 3 (h^1)^2 - 6 h^0 h^1\right) + \mathcal{O} (r_0^{-1})\right]\, ,
\end{align}
which cancels exactly the divergnce \eqref{eq:divergence} depending on the cutoff $r_0$ in the action \eqref{eq:ActionOnLocus}.

However, like in the $\textrm{AdS}_4$ case, the coupling between the boundary curvatures and the hypermultiplet introduces new divergence depending on $r_0$:
\be\label{eq:Sprime}
  S' = \int d^3 x\, \sqrt{g_3} (-A^2) \left(\frac{\kappa}{3} - \frac{L \mathcal{R}}{3} \right) = -\frac{\Omega_3^{\textrm{reg}} L^2}{4 \pi G} r_0^3 + \mathcal{O} (r_0^{-1})\, ,
\ee
where we have used $A^2 = 1 / (4 \pi G)$. To cancel this divergence, we have to take into account a boundary term for the flux:
\be\label{eq:Sflux}
  S_{\textrm{flux}} = -i \frac{N}{3 \pi^2} \int_{\textrm{S}^1 \times \mathbb{H}^2} C_3\, ,
\ee
where $C_3$ has the following form:
\be
  C_3 = \frac{a}{3} (r^3 - 1) d\Omega_3
\ee
with a constant $a$, such that $C_3$ vanishes at the horizon and the field strength $F_4 = dC_3$ satisfies
\be
  F_4 = a\, \omega_{\textrm{AdS}_4}\, ,
\ee
where $\omega_{\textrm{AdS}_4}$ is the volume form of the $\textrm{AdS}_4$ neutral topological black hole in our case. As discussed in Ref.~\cite{ABJM}, for ABJM theory, $X_7 = \textrm{S}^7 / \mathbb{Z}_k$, which is a Hopf fibration over $M_6 = \mathbb{C} \mathbb{P}^3$. Integrating $*F_4$ over $X_7$ gives
\be\label{eq:flux}
  \int_{X_7} *F_4 = 6 i L^6\, \textrm{Vol} (X_7)\, ,
\ee
which can also be related to the flux $N$ through \cite{ABJM, SenMarino}
\be
  N = \frac{6 L^6 \, \textrm{Vol} (X_7)}{(2 \pi \ell_P)^6}\, .
\ee
The condition \eqref{eq:flux} consequently fixes the constant $a = 3 i L^3 / 8$.

From the dimensional reduction of the 11-dimensional M-theory to 4 dimensions, we obtain
\be
  \frac{L^7\, \textrm{Vol} (X_7)}{64 \pi G_{11}} = \frac{1}{16 \pi G_4}\, ,
\ee
where $16 \pi G_{11} = (2 \pi)^8$ in the unit $\ell_P = 1$. Moreover, using the relation $\textrm{Vol} (X_7) \sim \textrm{Vol} (M_6) / k$ as well as the relation between $N$ and $\textrm{Vol} (X_7)$ discussed above, one can express $\Omega_3^{\textrm{reg}} L^2 / (4 \pi G_4)$ in terms of $k$ and $N$. After choosing an appropriate normalization factor, we have
\be
  -\frac{\Omega_3^{\textrm{reg}} L^2}{8 \pi G_4} = \frac{\sqrt{2} \pi}{3} k^{1/2} N^{3/2}\, .
\ee

Altogether, the divergence appearing in $S'$ \eqref{eq:Sprime} is canceled by the flux term $S_{\textrm{flux}}$ \eqref{eq:Sflux}, and a finite contribution from $S_{\textrm{flux}}$ remains:
\be
  S_{\textrm{flux}} \supset - \frac{\Omega_3^{\textrm{reg}} L^2}{4 \pi G_4} = \frac{2 \sqrt{2} \pi}{3} k^{1/2} N^{3/2}\, .
\ee
Finally, after the holographic renormalization the remaining finite part of the action is
\begin{align}
  S_{\textrm{finte}} & = \frac{\sqrt{2} \pi}{6} k^{1/2} N^{3/2} \Bigg[(h^1 - i)^{3/2} (h^0 - i)^{1/2} + i (h^0 + 3 h^1) + 4 \nonumber\\
  {} & \quad - 3 (h^1 + i)^{3/2} (h^0 + i)^{1/2} + 3 i (h^1 + i)^{1/2} (h^0 + i)^{1/2} + i (h^1 + i)^{3/2} (h^0 + i)^{-1/2} \Bigg] \nonumber\\
  {} & = \frac{\sqrt{2} \pi}{6} k^{1/2} N^{3/2} \Bigg[- (1 + i h^1)^{3/2} (1 + i h^0)^{1/2} + i (h^0 + 3 h^1) + 4 \nonumber\\
  {} & \quad + 3 (1 - i h^1)^{3/2} (1 - i h^0)^{1/2} - 3 (1 - i h^1)^{1/2} (1 - i h^0)^{1/2} - (1 - i h^1)^{3/2} (1 - i h^0)^{-1/2} \Bigg] \, .\label{eq:Sfinite}
\end{align}
As a check, we can also turn off all the fluctuations $h^0$ and $h^1$ in Eq.~\eqref{eq:Sfinite}, which gives us
\be
  S_{\textrm{finite}} (h^0 = 0, h^1 = 0) = - \frac{\Omega_3^{\textrm{reg}} L^2}{8 \pi G_4} = \frac{\pi L^2}{2 G_4} = \frac{\sqrt{2} \pi}{3} k^{1/2} N^{3/2}\, ,\label{eq:Sonshell}
\ee
where we have used the regularized volume for $\textrm{S}^1 \times \mathbb{H}^2$ \cite{RegVol-1, RegVol-2}, $\Omega_3^{\textrm{reg}} = - 4 \pi^2$. This expression of the finite action is exactly equal to the on-shell action of the $\textrm{AdS}_4$ neutral topological black hole \cite{Rey, Nishioka-2}.

\subsubsection{Evaluation of the Integral}

Finally, let us evaluate the path integral with the finte part of the action $S_{\textrm{finte}}$ after holographic renormalization given by Eq.~\eqref{eq:Sfinite}. Unlike the $\textrm{AdS}_4$ case discussed in Ref.~\cite{4DSUGRAloc}, the result of the path integral in this case is not an Airy function. Instead, we can apply the steepest descent method (see e.g. \cite{WittenCS}) to obtain the asymptotic expression, which suffices for our purpose of computing the entanglement entropy in the large-$N$ expansion.

We first find that $S_{\textrm{finte}}$ in Eq.~\eqref{eq:Sfinite} has only one critical point $(h^0,\, h^1) = (0,\, 0)$, hence there are no Stokes phenomena in this case. $S_{\textrm{finite}}$ evaluated at this critical point is equal to the on-shell value shown above in Eq.~\eqref{eq:Sonshell}. Next, we can expand $S_{\textrm{finite}}$ around this critical pont and perform the integration over $h^{0, 1}$ to obtain the asymptotic expression for the partition function $Z$, which is given by
\be\label{eq:pathintegral}
  Z = \int Dh^0\, Dh^1\, e^{- S_{\textrm{finite}}}\, .
\ee
In principle, there can also be some nontrivial Jacobian for $h^{0, 1}$ in the path integral.

Around the critical point $(h^0,\, h^1) = (0,\, 0)$, for $h^0 \sim \epsilon,\, h^1 \sim \epsilon$ the expansion to the order $\mathcal{O} (\epsilon^2)$ is given by
\be
  S_{\textrm{finite}} = \frac{\sqrt{2} \pi}{6} k^{1/2} N^{3/2} \left[2 + \frac{1}{4} (h^0)^2 - \frac{3}{2} h^0 h^1- \frac{3}{4} (h^1)^2 + o (\epsilon^2) \right]\, ,\label{eq:Sexp-1}
\ee
where the constant $2$ in the brackets gives the on-shell contribution. Let us compare this expansion with the one for the $\textrm{AdS}_4$ case. For the $\textrm{AdS}_4$ case discussed in Ref.~\cite{4DSUGRAloc}, the finite part of the action after holographic renormalization is given by
\be
  S_{\textrm{ren}} = -\frac{\pi \sqrt{2}}{3} k^{1/2} N^{3/2} \left[(1 - i h^1)^{3/2} \sqrt{1 - i h^0} + \frac{i}{2} (3 h^1 + h^0) - 2 \right]\, .
\ee
It has also only one critical point $(h^0,\, h^1) = (0,\, 0)$. We can similarly expand $S_{\textrm{ren}}$ around its critical point and obtain the expansion
\be
  S_{\textrm{ren}} = -\frac{\sqrt{2} \pi}{3} k^{1/2} N^{3/2} \left[-1 + \frac{1}{8} (h^0)^2 - \frac{3}{4} h^0 h^1 - \frac{3}{8} (h^1)^2 + o (\epsilon^2) \right]\, .\label{eq:Sexp-2}
\ee
We see that the expansion \eqref{eq:Sexp-2} for the $\textrm{AdS}_4$ case has exactly the same on-shell contribution as the expansion \eqref{eq:Sexp-1} for the topological black hole case, while at the order $\mathcal{O} (\epsilon^2)$ they only differ by a sign, which can be compensated by rotating the contours in the integrals. The true discrepancy takes place at higher orders $\sim o (\epsilon^2)$, which is not just a sign difference.

Although the partition function for the $\textrm{AdS}_4$ neutral topological black hole is not exactly given by an Airy function as in the $\textrm{AdS}_4$ case, based on the comparison above, we expect that the asymptotic expressions of the partition function for both cases should coincide at leading orders up to a phase factor. However, assuming a flat measure for $h^{0, 1}$ in the path integral \eqref{eq:pathintegral}, a direct computation of the partition function $Z$ by integrating $h^{0, 1}$ in $S_{\textrm{finite}}$ \eqref{eq:Sexp-1} does not give the result that we expect.
% Instead of $z^{1/4}$ in the denominator we would have a factor $z^{3/2}$ in the denominator.
Instead, if we adopt the assumptions made in Ref.~\cite{4DSUGRAloc} and perform the same procedure by dropping some constants from the Jacobian and the Gaussian integral, due to the same expansion of the renormalized action around the critical point, for the $\textrm{AdS}_4$ neutral topological black hole the partition function of the supergravity has the expected asymptotic expansion for $z \to \infty$:
\be\label{eq:Zasympt}
  Z \sim \frac{\textrm{exp} \left(-\frac{2}{3} z^{3/2} \right)}{2 \sqrt{\pi} (-z)^{1/4}} \left(1 + \mathcal{O} (z^{-3/2})  \right)\, ,
\ee
where $z \equiv N (\pi^2 k / 2)^{1/3}$.

Therefore, we find the partition function \eqref{eq:Zasympt} of the $4D$ $\mathcal{N}=2$ off-shell gauged supergravity on the background of the $\textrm{AdS}_4$ neutral topological black hole. Although this result is not exactly the same as the partition function for the ABJM theory on round $\textrm{S}^3$ \cite{Marino} or the partition function of the 4D $\mathcal{N}=2$ off-shell gauged supergravity on $\textrm{AdS}_4$ \cite{4DSUGRAloc}, which are equal to $\textrm{Ai}(z)$ perturbatively and have the asymptotic expansion for $-\pi < \textrm{arg} (z) < \pi$:
\be
  \textrm{Ai} (z) \sim \frac{e^{-\frac{2}{3} z^{3/2}}}{2 \sqrt{\pi} z^{1/4}} \left[\sum_{n=0}^\infty \frac{(-1)^n\, \Gamma\left(n + \frac{5}{6} \right)\, \Gamma\left(n + \frac{1}{6} \right)\, \left(\frac{3}{4} \right)^n}{2 \pi \, n!\, z^{3 n / 2}} \right]\, ,
\ee
all these theories share the same asymptotic expansion at the leading order up to a phase given by Eq.~\eqref{eq:Zasympt}, which consequently leads to the free energy:
\be
  F = - \textrm{log}\, Z = \frac{\sqrt{2} \pi}{3} k^{1/2} N^{3/2} + \frac{1}{4} \textrm{log}(N) + \mathcal{O} (N^0) \, ,
\ee
where we single out the logarithmic term, whose coefficient is universal and has some physical meaning that we will briefly discuss in the next section.

\section{Black Hole Entropy and Entanglement Entropy}\label{sec:Entropy}

After computing the partition function and the free energy of the 4D $\mathcal{N}=2$ off-shell gauged supergravity on the background of the $\textrm{AdS}_4$ neutral topological black hole, we can relate them to the black hole entropy according to Refs.~\cite{Rey, Nishioka-2}:
\be\label{eq:SBH}
  S_{BH} = - F = \textrm{log}\, Z = - \frac{\sqrt{2} \pi}{3} k^{1/2} N^{3/2} - \frac{1}{4} \textrm{log}(N) + \mathcal{O} (N^0) \, .
\ee
The leading term in the expression above corresponds to the contribution from the Bekenstein-Hawking entropy formula of the black hole \cite{Hawking, Bekenstein-1, Bekenstein-2}, i.e. $A / (4 G)$, while the second term is the logarithmic correction. Therefore, the supergravity localization indeed provides a way of computing the logarithmic corrections of the black hole entropy.

Based on the AdS/CFT correspondence or more precisely the gravity dual of supersymmetric R\'enyi entropy discussed in Refs.~\cite{Rey, Nishioka-2}, we can interpret the entropy of the $\textrm{AdS}_4$ neutral topological black hole obtained above from the supergravity localization as the entanglement entropy of the ABJM theory across a circle $\textrm{S}^1$ on the boundary, i.e.,
\be\label{eq:SEEgrav}
   S_{EE}^{\textrm{ABJM}} = S_{BH} = - \frac{\sqrt{2} \pi}{3} k^{1/2} N^{3/2} - \frac{1}{4} \textrm{log}(N) + \mathcal{O} (N^0) \, .
\ee
This interpretation is also consistent with the Ryu-Takayanagi formula \cite{RT, MaldacenaRT} of the entanglement entropy for the boundary conformal field theory.

The entanglement entropy of the ABJM theory can also be obtained directly from the field theory side by evaluating the $\textrm{S}^3$-partition function of the ABJM theory \cite{Kapustin, Marino} and expanding it at large $N$:
\be\label{eq:SEEfield}
  S_{EE}^{\textrm{ABJM}} = - \frac{\sqrt{2} \pi}{3} k^{1/2} N^{3/2} + \frac{\sqrt{2} \pi (8 + k^2)}{48} k^{-1/2} N^{1/2} - \frac{1}{4} \textrm{log}(N) + \mathcal{O} (N^0) \, .
\ee
In particular, we can consider a special case $k=1$:
\be
  S_{EE,\, k=1}^{\textrm{ABJM}} = - \frac{\sqrt{2} \pi}{3} N^{3/2} + \frac{3 \sqrt{2} \pi}{16} N^{1/2} - \frac{1}{4} \textrm{log}(N) + \mathcal{O} (N^0) \, .
\ee
This result also coincides with the $q\to 1$ limit of the supersymmetric R\'enyi entropy for $k=1$, which was discussed in  Refs.~\cite{Nishioka-1, Hatsuda}.

Comparing the results Eq.~\eqref{eq:SEEgrav} and Eq.~\eqref{eq:SEEfield}, we see that they differ in two places. First, the result from supergravity localization \eqref{eq:SEEgrav} does not have the contribution of the order $\sim N^{1/2}$. This is due to the fact that this term corresponds to the stringy corrections, which cannot be taken into account within supergravity, as discussed in Ref.~\cite{4DSUGRAloc}. More precisely, the resulf of supergravity localization on $\textrm{AdS}_4$ differs by a shift in $N$ compared to the one from field theory localization, i.e., instead of $N/k$ the field theory localization has
\be
  \frac{N}{k} \left(1 - \frac{1}{24} \frac{k}{N} - \frac{1}{3} \frac{1}{N k} \right) \sim \frac{N}{k} \left(1 + C_1 \frac{\ell^4}{L^4} + C_2\, g_4^2 \right)\, ,
\ee
where
\be
  \frac{N}{k} \sim \frac{L^4}{\ell^4}\, ,\quad \frac{1}{N k} \sim g_4^2\, ,
\ee
with $g_4$, $L$ and $g_4 \ell$ denoting the 4D string coupling, the $\textrm{AdS}_4$ radius and the 4D Planck length respectively. Hence, these shifts correspond to some stringy effects, and to reproduce them requires some stringy computation in the bulk. The similar thing happens in our calculations, i.e., the supergravity localization on the $\textrm{AdS}_4$ neutral topological black hole cannot reproduce some stringy corrections, which includes the term $\sim N^{1/2}$.

Second difference between Eq.~\eqref{eq:SEEgrav} and Eq.~\eqref{eq:SEEfield} is that the expression \eqref{eq:SEEfield} comes from the expansion of the Airy function, while Eq.~\eqref{eq:SEEgrav} does not. Although they have the same leading-order expression and the logarithmic correction, the discrepancy emerges at higher orders, since the result from supergravity localization \eqref{eq:SEEgrav} does not exactly reproduce an Airy function. This fact is closely related to the recent works of field theory localization on noncompact manifolds \cite{ICTP, Martelli, noncptIndex}, whose results depend on the boundary conditions. In particular, Ref.~\cite{noncptIndex} has discussed the localization of the ABJM theory on $\textrm{S}^1 \times \mathbb{H}^2$, which is precisely the field theory dual of the gravity considered in this paper. Instead of the untwisted partition function, in Ref.~\cite{noncptIndex} the authors focused on the topologically twisted index of the ABJM theory on $\textrm{S}^1 \times \mathbb{H}^2$ similar to Ref.~\cite{BeniniGrav} and evaluated it at the leading order in $N$. We expect that the large-$N$ result of the topologically twisted index in a special limit could match exactly our result obtained from the gravity side, which requires an analysis of the matrix integral beyond the leading order of $N$.

Concerning the black hole entropy, there have already been some works devoted to this topic using the supergravity localization technique (e.g. \cite{BH2010, BH2011, Joao13, BHloc-1, BHloc-2, Joao15}). Our work provides a concrete example and relates the bulk black hole entropy to the entanglement entropy of the boundary conformal field theory. In particular, the near-horizon geometry of the $\textrm{AdS}_4$ neutral topological black hole \eqref{eq:AdS4metric} that we discussed is $\textrm{AdS}_2 \times \mathbb{H}^2$. According to Ref.~\cite{BHloc-1}, the quantum black hole entropy has the general form:
\be\label{eq:SBHgeneral}
  S_{\textrm{BH}}^{\textrm{qu}} = \frac{A}{4} + a_0 \, \textrm{log} A + \cdots\, ,
\ee
where $a_0$ is the coefficient that can be computed by the supergravity localization for each multiplet on the black hole background or by direct evaluation of the 1-loop determinant around the classical attractor background \cite{Sen-1, Sen-2, Sen-3, Sen-4, Sen-5, Lisbao, Liu, JeonLal}. We can compare the general expression \eqref{eq:SBHgeneral} with our result \eqref{eq:SBH}. The identification of the leading terms implies that
\be
  A \sim N^{3/2}\, .
\ee
Together with the identification of the logarithmic terms, we deduce that in our case
\be
  a_0 = -\frac{1}{6}\, ,
\ee
which is twice the contribution of one vector multiplet $a_0^{\textrm{vec}} = -1/12$ \cite{BHloc-1}. This is consistent with the fact in our case only $(n_v + 1) = 2$ vector multiplets have fluctuations, while the metric and the hypermultiplet are fixed to be the attractor solutions.

Moreover, the logarithmic correction to the black hole entropy was also obtained on-shell from the 1-loop computation in the Euclidean 11-dimensional supergravity on $\textrm{AdS}_4 \times X_7$, and the result coincides with the logarithmic term in the large-$N$ expansion of the partition function of the ABJM theory on $\textrm{S}^3$ \cite{SenMarino}, which is also consistent with our result from the localization of the 4D $\mathcal{N}=2$ off-shell supergravity on the $\textrm{AdS}_4$ neutral topological black hole.

%%%%%%%%%%%%%%%%%%%%%%%%%%%%%%%%%%%%%%%%%%%
%%%%%%%%%%%%%%%%%%%%%%%%%%%%%%%%%%%%%%%%%%%
\section{Conclusion and Discussion}\label{sec:discussion}
%%%%%%%%%%%%%%%%%%%%%%%%%%%%%%%%%%%%%%%%%%%
%%%%%%%%%%%%%%%%%%%%%%%%%%%%%%%%%%%%%%%%%%%

In this paper we have calculated the partition function of the 4D $\mathcal{N}=2$ off-shell gauged supergravity in the background of the $\textrm{AdS}_4$ neutral topological black hole via supersymmetric localization. The free energy of the theory is related to the black hole entropy, and using the localization we obtain the logarithmic correction to the leading order result given by the Bekenstein-Hawking formula. Moreover, we compare the black hole entropy with the entanglement entropy of the ABJM theory across a circle $\textrm{S}^1$ on the boundary and find an exact match up to some stringy effects, which provides a precise test of the AdS/CFT correspondence beyond the leading order.

There are many more interesting extensions of this work for the future research. For instance, one can compute supersymmetric R\'enyi entropy of the boundary ABJM theory via supergravity localization on backgrounds of charged topological black holes, which generalizes the classical results of Refs.~\cite{Rey, Nishioka-2} on the gravity side and can also be compared with the exact results on the field theory side discussed in Ref.~\cite{Hatsuda}. Another possible extension is to compute the supersymmetric Wilson loop via supergravity localization. The result can be compared with the exact result on the field theory side \cite{Kapustin}, and also generalizes the classical result of Ref.~\cite{Nishioka-2}.

Related to the recent works on the supersymmetric localization of field theories on noncompact manifolds \cite{Martelli, ICTP} and in particular the topologically twisted index of the ABJM theory on $\textrm{S}^1 \times \mathbb{H}^2$ discussed in Ref.~\cite{noncptIndex}, we expect that the localization result from the field theory side can reproduce our result obtained from the supergravity side beyond the leading order in $N$. In general, the study of the field theory localization on noncompact manifolds is very important, not only because it provides another precise test of the AdS/CFT correspondence, also because it is directly related to some exact computations of entanglement entropy, supersymmetric R\'enyi entropy as well as the bulk black hole entropy.

Recently, in Ref.~\cite{BeniniGrav} Benini, Hristov and Zaffaroni have found a new relation between the topologically twisted index of the ABJM theory on $\textrm{S}^2 \times \textrm{S}^1$ and the entropy of the 4D STU black hole, which in principle allows one to count the microstates of the black hole in the dual field theory. As we mentioned earlier, this work has also been generalized to the ABJM theory on $S^1 \times \mathbb{H}^2$ and correspondingly the $\textrm{AdS}_4$ hyperbolic black hole \cite{noncptIndex}. Using the technique of localization of supergravity to the near-horizon geometry of the 4D STU black hole, we should be able to test this new correspondence beyond the leading order and compare the results also to the supersymmetric R\'enyi entropy.

Finally, supergravity localization itself still needs more study. As we have seen from the text, the metric is fixed as a background, i.e., we have not taken into account the fluctuations of the Weyl multiplet. Although some indirect result can be deduced \cite{BHloc-1}, as far as we know, there is still no direct computation of the localization of the Weyl multiplet in the literature. In some sense, we are studying supergravity as a special kind of quantum field theory on a curved manifold, which shares the same spirit of the work by Festuccia and Seiberg \cite{SeibergFestuccia}. More detailed study is definitedly required to truly understand the behavior of supergravity. We would like to investigate this open problem in the future research.

\section*{Acknowledgements}

We would like to thank Chris Elliott, Rouven Frassek, Jo\~ao Gomes, Song He, Chris Herzog, Sungjay Lee, Hai Lin, Satoshi Nawata, Nikita Nekrasov, Vasily Pestun, Naveen Prabhakar, Jianfei Xu, Itamar Yaakov, Peng Zhao and Yang Zhou for many useful discussions. The work of XZ was supported in part by the NSF grant PHY 1404446.

\appendix
\section{Review of 4D $\mathcal{N}=2$ Off-Shell Gauged Supergravity}\label{app:SUGRA}

We review the 4D $\mathcal{N}=2$ off-shell gauged supergravity theory in this appendix. It can be obtained as a consistent truncation of M-theory on a Sasaki-Einstein manifold $X_7$. The theory was originally constructed in Ref.~\cite{deWit} and also reviwed in Ref.~\cite{4DSUGRAloc}. We follow these references closely.

The nearly massless fields consist of the supergravity multiplet, a single vector multiplet and a universal hypermultiplet (the dualized tensor multiplet). To obtain an off-shell super-Poincar\'e gravity theory, one can start with a superconformal gravity theory and then use gauge fixing to reduce it to the super-Poincar\'e gravity theory.

The $\mathcal{N}=2$ superconformal algebra has the generators:
\be
  P_a,\, M_{ab},\, D,\, K_a,\, Q_i,\, S^i,\, U_{ij},
\ee
where $\{P_a,\, M_{ab},\, D,\, K_a \}$ are the generators of translations, Lorentz rotations, dilatations, special conformal transformations respectively, whlie $Q_i$ and $S^i$ are the usual supersymmetry and the special conformal supersymmetry generator respectively, and $U_{ij}$ is the generator of the 4D $SU(2)$ R-symmetry. The gauge fields corresponding to these generators are
\be
  e_\mu^a,\, \omega_\mu^{ab},\, b_\mu,\, f_\mu^a,\, \psi_\mu^i,\, \phi_\mu^i,\, \mathcal{V}_\mu^{ij}
\ee
respectively.

Let us review different $\mathcal{N}=2$ supersymmetry multiplets in the following:

\begin{itemize}
\item Weyl multiplet:

The Weyl multiplet, denoted by $\bold{W}$, contains the following field components:
\be
  \bold{W} = \big( e_\mu^a,\, \psi_\mu^i,\, b_\mu,\, A_\mu,\, \mathcal{V}_{\mu j}^i,\, T_{ab}^{ij},\, \chi^i,\, D \big)\, ,
\ee
where $e_\mu^a$ is the vielbein, $\psi_\mu^i$ is the (left-handed) gravitino doublet, $b_\mu$ and $A_\mu$ are the gauge fields of dilatations and chiral $U(1)$ R-symmetry transformations repectively, while $\mathcal{V}_{\mu j}^i$ is the gauge field of the $SU(2)$ R-symmetry. The auxiliary fields include the antisymmetric anti-selfdual field $T_{ab}^{ij}$, the $SU(2)$ doublet Majorana spinor $\chi_i$ and the real scalar $D$. Altogether, there are 24 bosonic degrees of freedom and 24 fermionic degrees of freedom in the Weyl multiplet. They satisfy the following supersymmetric transformations:
\begin{align}
  \delta e_\mu\,^a & = \bar{\epsilon}^i \gamma^a \psi_{\mu i} + \bar{\epsilon}_i \gamma^a \psi_\mu\,^i\, ,\nonumber\\
  \delta \psi_\mu\,^i & = 2\, \mathcal{D}_\mu \epsilon^i - \frac{1}{8} T_{ab}\,^{ij} \gamma^{ab} \gamma_\mu \epsilon_j - \gamma_\mu \eta^i\, ,\nonumber\\
  \delta b_\mu & = \frac{1}{2} \bar{\epsilon}^i \phi_{\mu i} - \frac{3}{4} \bar{\epsilon}^i \gamma_\mu \chi_i - \frac{1}{2} \bar{\eta}^i \psi_{\mu i} + \textrm{h.c.} + \Lambda_K^a e_{\mu a}\, ,\nonumber\\
  \delta A_\mu & = \frac{i}{2} \bar{\epsilon}^i \phi_{\mu i} + \frac{3 i}{4} \bar{\epsilon}^i \gamma_\mu \chi_i + \frac{i}{2} \bar{\eta}^i \psi_{\mu i} + \textrm{h.c.}\, ,\nonumber\\
  \delta \mathcal{V}_\mu\,^i\,_j & = 2 \bar{\epsilon}_j \phi_\mu\,^i - 3 \bar{\epsilon}_j \gamma_\mu \chi^i + 2 \bar{\eta}_j \psi_\mu\,^i - (\textrm{h.c.})\, ,\nonumber\\
  \delta T_{ab}\,^{ij} & = 8 \bar{\epsilon}^{[i} R(Q)_{ab}\,^{j]}\, ,\nonumber\\
  \delta \chi_i & = -\frac{1}{12} \gamma^{ab} \slashed D T_{ab}\,^{ij} \epsilon_j + \frac{1}{6} R(\mathcal{V})_{\mu\nu}\,^i\,_j \gamma^{\mu\nu} \epsilon^j - \frac{i}{3} R_{\mu\nu}(A) \gamma^{\mu\nu} \epsilon^i + D \epsilon^i + \frac{1}{12} \gamma_{ab} T^{abij} \eta_j\, ,\nonumber\\
  \delta D & = \bar{\epsilon}^i \slashed D \chi_i + \bar{\epsilon}_i \slashed D \chi^i\, ,
\end{align}
where $\epsilon$, $\eta$ and $\Lambda_K^a$ denote the parameters of $Q$, $S$ and $K_a$ respectively, and
\be
  \mathcal{D}_\mu \epsilon^i \equiv \left(\partial_\mu + \frac{1}{4} \omega_\mu\,^{cd} \gamma_{cd} + \frac{1}{2} b_\mu + \frac{i}{2} A_\mu \right) \epsilon^i + \frac{1}{2} \mathcal{V}_\mu\,^i\,_j \epsilon^j\, .
\ee

\item Vector multiplet:

The vector multiplet, denoted by $\bold{X}^I$ with the index $I$ labelling the gauge group generators, contains the following field components:
\be
  \bold{X}^I = \big( X^I,\, \Omega_i^I,\, W_\mu^I,\, Y_{ij}^I \big)\, ,
\ee
where $X^I$ is a complex scalar, $\Omega_i^I$ is the gaugino that are the $SU(2)$ doublet of chiral fermions, and $W_\mu^I$ is the vector field. The auxiliary field $Y_{ij}^I$ is an $SU(2)$ triplet with
\be
  Y_{ij} = Y_{ji}\, ,\quad Y_{ij} = \epsilon_{ik}\, \epsilon_{jl}\, Y^{kl}\, .
\ee
Altogether, there are 8 bosonic degrees of freedom and 8 fermionic degrees of freedom in the vector multiplet for each index $I$. They have the following supersymmetric transformations:
\begin{align}
  \delta X & = \bar{\epsilon}^i \Omega_i\, ,\nonumber\\
  \delta \Omega_i & = 2 \slashed D X \epsilon_i + \frac{1}{2} \epsilon_{ij} \mathcal{F}_{\mu\nu} \gamma^{\mu\nu} \epsilon^j + Y_{ij} \epsilon^j + 2 X \eta_i\, ,\nonumber\\
  \delta W_\mu & = \epsilon^{ij} \bar{\epsilon}_i (\gamma_\mu \Omega_j + 2 \psi_{\mu j} X) + \epsilon_{ij} \bar{\epsilon}^i (\gamma_\mu \Omega^j + 2 \psi_\mu\,^j \overline{X})\, ,\nonumber\\
  \delta Y_{ij} & = 2 \bar{\epsilon}_{(i} \slashed D \Omega_{j)} + 2 \epsilon_{ij} \epsilon_{jl} \bar{\epsilon}^{(k} \slashed D \Omega^{l)}\, ,
\end{align}
where
\be
  \mathcal{F}_{\mu\nu} \equiv F_{\mu\nu} - \frac{1}{4} \left(\overline{X} \epsilon_{ij} T_{\mu\nu}\,^{ij} + \textrm{h.c.} \right) + (\textrm{fermionic terms})\, ,
\ee
and $F_{\mu\nu} = \partial_\mu W_\nu - \partial_\nu W_\mu$.

\item Hypermultiplet:

The hypermultiplet of the 4D $\mathcal{N}=2$ supersymmetry is a little special, because it is well-known that for this multiplet the off-shell closure of the supersymmetry algebra cannot be achieved with finite number of fields. One can start from one hypermultiplet, and then add infinite sequence of fields to obtain the off-shell closure of the supersymmetry algebra.

A single hypermultiplet contains scalars $A_i\,^\alpha$ and spinors $\zeta^\alpha$, where the scalars are doublets of the $SU(2)$ R-symmetry, and all the fields transform in the fundamental representation of $Sp(2r)$, i.e. the index $\alpha$ runs from $1$ to $2r$. Moreover, the scalars satisfy the reality condition
\be
  A^i_\alpha = (A_i^\alpha)^* = \epsilon^{ij}\, \rho_{\alpha\beta}\, A_j^\beta
\ee
with
\be
  \rho_{\alpha\beta}\, \rho^{\beta\gamma} = - \delta_\alpha^\gamma\, .
\ee
The supersymmetric transformations are given by
\begin{align}
  \delta A_i\,^\alpha & = 2 \bar{\epsilon}_i \zeta^\alpha + 2 \rho^{\alpha\beta} \epsilon_{ij} \bar{\epsilon}^j \zeta_\beta\, ,\nonumber\\
  \delta \zeta^\alpha & = \slashed D A_i\,^\alpha \epsilon^i + 2 g X^\alpha\,_\beta A_i\,^\beta \epsilon^{ij} \epsilon_j + A_i\,^\alpha \eta^i\, ,\label{eq:SUSYforhyper}
\end{align}
where $g$ is the coupling constant,
\be
  X^\alpha\,_\beta \equiv X^I (t_I)^\alpha\,_\beta\, ,\quad \overline{X}^\alpha\,_\beta \equiv \overline{X}^I (t_I)^\alpha\,_\beta
\ee
with $t_\alpha\,^\beta \rho_{\beta\gamma} = \rho_{\alpha\beta} t^\beta\,_\gamma$, and
\be
  D_\mu A_i\,^\alpha \equiv \partial_\mu A_i\,^\alpha + \frac{1}{2} \mathcal{V}_\mu^j\,_i A_j\,^\alpha - b_\mu A_i\,^\alpha - g W^\alpha_{\mu\beta} A_i\,^\beta - \bar{\psi}_{\mu i} \zeta^\alpha - \rho^{\alpha\beta} \epsilon_{ij} \bar{\psi}_\mu^j \zeta_\beta\, .
\ee

As discussed in Refs.~\cite{deWit, 4DSUGRAloc}, to realized the off-shell supersymmetry for the hypermultiplet, one needs to introduce an infinite tower of hypermultiplets $(A_i\,^\alpha,\, \zeta^\alpha)$, $(A_i\,^\alpha,\, \zeta^\alpha)^{(z)}$, $(A_i\,^\alpha,\, \zeta^\alpha)^{(zz)}$, $\cdots$. The closure of the superconformal algebra will then impose an infinite set of constraints, and in the end only $(A_i\,^\alpha,\, \zeta^\alpha, A_i\,^{\alpha (z)})$ are independent. Consequently, the supersymmetric transformation \eqref{eq:SUSYforhyper} for $\zeta^\alpha$ will be modified to
\be
  \delta \zeta^\alpha = \slashed D A_i\,^\alpha \epsilon^i + 2 g X^\alpha\,_\beta A_i\,^\beta \epsilon^{ij} \epsilon_j + A_i\,^\alpha \eta^i + a A_i\,^{a(z)} \epsilon^{ij} \epsilon_j\, ,
\ee
where $a$ is the scalar field in the vector multiplet associated with the central charge translation, which can be set to $a=1$. In the main text, we also denote $A_i\,^{\alpha (z)}$ by $F_i\,^\alpha$ using
\be
  F_i\,^\alpha = a A_i\,^{\alpha (z)}\, .
\ee

After constructing a linear multiplet coupled to $(A_i\,^\alpha,\, \zeta^\alpha)$, $(A_i\,^\alpha,\, \zeta^\alpha)^{(z)}$ and imposing the constraints, one can find the supersymmetric Lagrangian for the hypermultiplet \cite{deWit, 4DSUGRAloc}:
\begin{align}
  \mathcal{L}_{\textrm{hyp}} & = \bigg[ - D_\mu A^i\,_\beta D^\mu A_i\,^\alpha - \frac{1}{6} R A^i\,_\beta A_i\,^\alpha + \frac{1}{2} D A^i\,_\beta A_i\,^\alpha + (|a|^2 + W_\mu^z W^{\mu z}) A^i\,_\beta\,^{(z)} A_i\,^{\alpha (z)} \nonumber\\
  {} & \quad + 4 g^2 A^i\,_\beta \overline{X}^\alpha\,_\gamma X^\gamma\,_\delta A_i\,^\delta + g A^i\,_\beta (Y^{ij})_\gamma\,^\alpha A_k\,^\gamma \epsilon_{ij} \bigg] d_\alpha\,^\beta + (\textrm{fermionic terms})\, ,
\end{align}
where $d_\alpha\,^\beta$ satisfies
\be
  \overline{d_\alpha\,^\beta} = d_\beta\,^\alpha\, ,
\ee
\be
  d_\alpha\,^\beta = \epsilon_{\gamma\alpha} \epsilon^{\delta\beta} d_\delta\,^\gamma\, ,
\ee
\be
  t^\alpha\,_\alpha d_\gamma\,^\beta + d_\alpha\,^\gamma t_\gamma\,^\beta = 0\, .
\ee
As discussed in Refs.~\cite{deWit, 4DSUGRAloc}, one can set $d_\alpha\,^\beta = - \delta_\alpha\,^\beta$.

\end{itemize}

\section{Killing Spinors and Gamma Matrices}\label{app:Killing}

To localize the 4D $\mathcal{N}=2$ off-shell gauged supergravity on the neutral topological black hole, we need to find the Killing spinors in this space. They are explicitly constructed in Ref.~\cite{Rey}, and we review the results in this and next appendix.

The metric of the Euclidean $\textrm{AdS}_4$ topological black hole is
\be\label{eq:EuclideanAdS4metric}
  ds^2 = f(r)\, d\tau^2 + \frac{1}{f(r)}\, dr^2 + r^2 \, d\Sigma(\mathbb{H}^2) = f(r)\, d\tau^2 + \frac{1}{f(r)}\, dr^2 + r^2 \, (du^2 + \textrm{sinh}^2 u\, d\phi^2)\, ,
\ee
where
\be
  f(r) = \frac{r^2}{L^2} + \kappa - \frac{2m}{r} + \frac{Q^2}{r^2}\, ,
\ee
and $2 \kappa$ is the constant curvature of the 2-dimensional Riemann surface, which implies that $\kappa = -1$ for $\mathbb{H}^2$.

Near the boundary ($r \to \infty$), we keep the terms $\sim \mathcal{O} (1)$ in $f(r)$, and the metric approaches
\be
  ds^2 = d\eta^2 + \textrm{sinh}^2 \eta\, d\tau^2 + \textrm{cosh}^2 \eta (du^2 + \textrm{sinh}^2 u\, d\phi^2)\, .
\ee
where
\be
  r = \textrm{cosh}\, \eta\, ,
\ee
and for simplicity we set $L=1$. The ranges of the variables are
\be
  \tau \in [0,\, 2 \pi q)\, ,\quad \eta \in [0,\, \infty)\, ,\quad \phi \in [0,\, 2 \pi)\, .
\ee
Hence, at $r \to \infty$ or $\eta \to \infty$ the boundary of the Euclidean $\textrm{AdS}_4$ topological black hole is $\textrm{S}^1 \times \mathbb{H}^2$ as expected.

In this paper, we adopt the convention of the $\gamma$-matrices used in Ref.~\cite{Rey}. For the Lorentz signature:
\be
  \gamma_0 =
  \left( \begin{array}{cccc}
    0 & 0 & 0 & -1\\
    0 & 0 & 1 & 0\\
    0 & -1 & 0 & 0\\
    1 & 0 & 0 & 0
  \end{array} \right)\, ,\quad
  \gamma_1 =
  \left( \begin{array}{cccc}
    -1 & 0 & 0 & 0\\
    0 & 1 & 0 & 0\\
    0 & 0 & -1 & 0\\
    0 & 0 & 0 & 1
  \end{array} \right)\, ,
\ee
\be
  \gamma_2 =
  \left( \begin{array}{cccc}
    0 & 0 & 0 & -1\\
    0 & 0 & 1 & 0\\
    0 & 1 & 0 & 0\\
    -1 & 0 & 0 & 0
  \end{array} \right)\, ,\quad
  \gamma_3 =
  \left( \begin{array}{cccc}
    0 & 1 & 0 & 0\\
    1 & 0 & 0 & 0\\
    0 & 0 & 0 & 1\\
    0 & 0 & 1 & 0
  \end{array} \right)\, .
\ee
For the Euclidean signature, one can choose $\gamma_4 = - i \gamma_0$ with the same $\gamma_i$'s. The charge conjugation matrix $B$ satisfies
\be
  B^{-1} \gamma_\mu B = - \gamma_\mu^*\, ,\quad B B^* = - \mathbb{I}\, .
\ee
More explicitly, in this paper we choose $B = \gamma_0$.

The Killing spinor equation for the $\textrm{AdS}_4$ topological black hole is \cite{Rey}
\be
  \nabla_\mu \epsilon - i g A_\mu \epsilon + \frac{i}{4} F_{ab} \gamma^{ab} \gamma_\mu \epsilon = -\frac{1}{2} g \gamma_\mu \epsilon\, .
\ee
Using the charge conjugation matrix $B$, one can construct the charge conjugate spinor $\epsilon^c \equiv B \epsilon^*$, which satisfies a different Killing spinor equation:
\be
  \nabla_\mu \epsilon^c + i g A_\mu \epsilon^c + \frac{i}{4} F_{ab} \gamma^{ab} \gamma_\mu \epsilon^c = \frac{1}{2} g \gamma_\mu \epsilon^c\, .
\ee

\section{Localization Action}\label{app:LocAction}

As discussed in Subsection~\ref{sec:LocAction}, we choose the Killing spinor found in Ref.~\cite{Rey}:
\be
  \epsilon = e^{-\frac{i}{2 q L} \tau_E}\, e^{i \frac{u}{2} \gamma_4 \gamma_1 \gamma_2}\, e^{\frac{\phi}{2} \gamma_{23}}\, \tilde{\epsilon}(r)
\ee
with
\be
  \tilde{\epsilon}(r) = \left(\sqrt{\frac{r}{L} + \sqrt{f(r)}} - i \gamma_4 \sqrt{\frac{r}{L} - \sqrt{f(r)}} \right) \left(\frac{1 - \gamma_1}{2} \right) \epsilon'_0\, ,
\ee
where $\epsilon'_0$ is an arbitrary constant spinor, and in this paper we choose $\epsilon'_0 = (1,\, 0,\, i,\, 0)^T$. We define
\be
  \xi^i \equiv \frac{1 + i \gamma_5}{2}\, \widetilde{\xi}^i
\ee
with
\be
  \widetilde{\xi}^1 \equiv \epsilon^c\, ,\quad \widetilde{\xi}^2 \equiv \epsilon\, .
\ee
and use $\xi^i = (\xi^1,\, \xi^2)^T$ as the Killing spinor in the expansion of the localization action.

With the Killing spinor chosen above, we can work out all the Killing spinor bilinears explicitly, and use them to expand the localization action.
%For instance,
%\begin{align}
%  \xi^\dagger \xi & =
%\end{align}
After some steps, we found that the localization action in this case can be written into a sum of some squares as follows:
\begin{align}
  (\delta \Omega)^\dagger \delta \Omega & = \frac{1}{2\, \textrm{cosh} (\eta + u)} \left(F_{ab}\, \textrm{cosh} (\eta + u) - \frac{1}{2} \epsilon_{abcd} F^{cd} - 2\, \epsilon_{abcd}\, \partial^c J\, V^d - 2\, \Theta_{ab}\, J \right)^2 \nonumber\\
  {} & \quad + \frac{1}{\textrm{cosh} (\eta + u)} \left(-\frac{1}{2} \epsilon_{abcd} F^{bc} V^d + 2 \partial_a \left(J\, \textrm{cosh} (\eta + u) \right) \right)^2 + \frac{4}{\textrm{cosh} (\eta + u)} (V^a \partial_a J)^2 \nonumber\\
  {} & \quad + \frac{1}{\textrm{cosh} (\eta + u)} \left(F_{ab} V^b - 2\, \partial_a J \right)^2 + 8\, \textrm{cosh} (\eta + u) \left(\partial_i H \right)^2 \nonumber\\
  {} & \quad + 8\, \textrm{cosh} (\eta + u) \left(\partial_\eta H + \frac{\textrm{sinh} (\eta + u)}{\textrm{cosh} (\eta + u)} H \right)^2 + 2\, \textrm{cosh} (\eta + u) \left(Y^1\,_1 - \frac{2 H}{\textrm{cosh} (\eta + u)} \right)^2 \nonumber\\
  {} & \quad + \frac{1}{2\, \textrm{cosh} (\eta - u)} \left(F_{ab}\, \textrm{cosh} (\eta - u) - \frac{1}{2} \epsilon_{abcd} F^{cd} - 2\, \epsilon_{abcd}\, \partial^c J\, V^d - 2\, \Theta_{ab}\, J \right)^2 \nonumber\\
  {} & \quad + \frac{1}{\textrm{cosh} (\eta - u)} \left(-\frac{1}{2} \epsilon_{abcd} F^{bc} V^d + 2 \partial_a \left(J\, \textrm{cosh} (\eta - u) \right) \right)^2 + \frac{4}{\textrm{cosh} (\eta - u)} (V^a \partial_a J)^2 \nonumber\\
  {} & \quad + \frac{1}{\textrm{cosh} (\eta - u)} \left(F_{ab} V^b - 2\, \partial_a J \right)^2 + 8\, \textrm{cosh} (\eta - u) \left(\partial_i H \right)^2 \nonumber\\
  {} & \quad + 8\, \textrm{cosh} (\eta - u) \left(\partial_\eta H + \frac{\textrm{sinh} (\eta - u)}{\textrm{cosh} (\eta - u)} H \right)^2 + 2\, \textrm{cosh} (\eta - u) \left(Y^1\,_1 - \frac{2 H}{\textrm{cosh} (\eta - u)} \right)^2\, ,\label{eq:SquareSum}
\end{align}
where $\eta$ is defined by $r = \textrm{cosh} (\eta)$, and $V_a$ and $\Theta_{ab}$ are defined as
\begin{align}
  V_a & = \frac{1}{4} (\xi^i)^\dagger \gamma_a \xi^i\, ,\\
  \Theta_{ab} & = -\frac{i}{4} (\xi^\dagger)^\dagger \gamma_{ab} \gamma_5 \eta^i\, .
\end{align}
By requiring all the squares in the sum of the localization action to vanish, we obtain the localization locus of the theory. We see that there is nonvanishing localization locus only at $u=0$:
\be
  H = \frac{C}{\textrm{cosh} (\eta)}\, ,\quad Y^1\,_1 = \frac{2 C}{\textrm{cosh}^2 (\eta)}\, ,
\ee
where $C$ is an arbitrary constant, and in Section~\ref{sec:SUGRAloc} we also use the parametrization $C = J h$. We make the gauge choice $A_t = 0$, and in this gauge the field $J$ and $F_{ab}$ satisfy
\be
  J = \textrm{const}\, ,\quad F_{ab} V^b = 0\, ,
\ee
where the constant value of $J$ is fixed by the attractor solutions \eqref{eq:attSol-1} \eqref{eq:attSol-2}.

For the hypermultiplet, as discussed in Appendix~\ref{app:SUGRA}, in principle we need an infinite tower $(A_i\,^\alpha,\, \zeta^\alpha)$, $(A_i\,^\alpha,\, \zeta^\alpha)^{(z)}$, $(A_i\,^\alpha,\, \zeta^\alpha)^{(zz)}$, $\cdots$ with constraints to realize the off-shell supersymmetry. To look for the BPS solutions, we should require
\be
  \delta \zeta^\alpha = 0\, ,\quad \delta \zeta^{\alpha (z)} = 0\, ,\quad \delta \zeta^{\alpha (zz)} = 0\, ,\quad \cdots
\ee
with respect to the constraints, which is rather involved. Instead we follow the approach applied in Ref.~\cite{4DSUGRAloc} by requiring $\delta \zeta = 0$ for all 8 Killing spinors, which consequently leads to the solutions
\be
  F_i\,^\alpha = - \frac{2 i g}{\sqrt{8 \pi G}} (\sigma_3)^\alpha\,_j (H\cdot P)\, ,\quad 2 g (J\cdot P) = - \frac{1}{L}
\ee
with $F_i\,^\alpha$ and $P_I$ given by
\be
  F_i\,^\alpha = a A_i\,^{\alpha (z)}\, ,\quad t_I A_i\,^\alpha = P_I (i \sigma_3)^\alpha\,_\beta A_i\,^\beta\, .
\ee
These solutions coincide with the solutions \eqref{eq:BPSsolhyper} to the BPS equations under the attractor solution \eqref{eq:attsolA}.

\section{Evaluation of the Action}\label{app:Action}

In this appendix, we evaluate the action \eqref{eq:SUGRAaction} along the localization locus found in Appendix~\ref{app:LocAction}. As we explained in Subsection~\ref{sec:Action}, up to the volume of the boundary manifolds, the integrals over the radial direction have the same integrand for the $\textrm{AdS}_4$ case discussed in Ref.~\cite{4DSUGRAloc} and the $\textrm{AdS}_4$ neutral topological black hole considered in this paper. However, the discrepancy comes from the measure $d^4 x\, \sqrt{g}$, which differs for the hyperbolic $\textrm{AdS}_4$ and the $\textrm{AdS}_4$ neutral topological black hole.

Let us briefly list the results in the following. For the choice of the prepotential $F(X) = \sqrt{X^0 (X^1)^3}$, one can compute the tensor $N_{IJ}$ defined by
\be
  N_{IJ} \equiv \frac{1}{2 i} \left(F_{IJ} - \overline{F}_{IJ} \right)\, ,\quad F_{IJ} \equiv \partial_I \partial_J F(X)\, .
\ee
The explicit expressions are
\be
  N_{00} = \frac{i}{8} \left(\frac{J^1}{J^0} \right)^{\frac{3}{2}} \left(t^3 + \bar{t}^3 \right)\, ,\quad N_{11} = - \frac{3 i}{8} \left(\frac{J^1}{J^0} \right)^{-\frac{1}{2}} \left(\frac{1}{t} + \frac{1}{\bar{t}} \right)\, ,
\ee
\be
  N_{01} = - N_{10} = -\frac{3 i}{8} \left(\frac{J^1}{J^0} \right)^{\frac{1}{2}} \left(t + \bar{t} \right)\, .
\ee
Plugging these expressions into $S_{\textrm{vec}}$, we obtain the following integral:
\begin{align}
  S_{\textrm{vec}} & = \Omega_3^{\textrm{reg}} L^2 \frac{J^0 (J^1)^3}{2 i} \int dr\, r^2 \Bigg[ - \left(1 + \frac{(h^0)^2}{r^2} \right) (t + \bar{t})^3 + \frac{3}{4} (h^1)^2 \left(\frac{1}{t} + \frac{1}{\bar{t}} \right) \frac{r^2 - 1}{r^4} \nonumber\\
  {} & \quad + \frac{3}{2} h^1 h^0 (t + \bar{t}) \frac{r^2 - 1}{r^4} - \frac{1}{4} (h^0)^2 \left(t^3 + \bar{t}^3 \right) \frac{r^2 - 1}{r^4} - \frac{3}{4} \left(\frac{1}{t} + \frac{1}{\bar{t}} \right) \left(1 + i \frac{h^1}{r^2} \right)^2 \nonumber\\
  {} & \quad - \frac{3}{2} (t + \bar{t}) \left(1 + i \frac{h^1}{r^2} \right) \left(1 + i \frac{h^0}{r^2} \right) + \frac{1}{4} \left(t^3 + \bar{t}^3 \right) \left(1 + i \frac{h^0}{r^2} \right)^2 \Bigg]\, ,
\end{align}
where $\Omega_3^{\textrm{reg}}$ is the regularized volume of the boundary $\textrm{S}^1 \times \mathbb{H}^2$. As we have seen in Appendix~\ref{app:LocAction}, the nontrivial localization locus is only supported by $u=0$, while the unregularized volume of the noncompact manifold $\mathbb{H}^2$ is divergent. Combining these two factors, we assume that the regularized volume $\Omega_3^{\textrm{reg}}$ of the boundary manifold $\textrm{S}^1 \times \mathbb{H}^2$ is finite.

The integral appearing in $S_{\textrm{vec}}$ can be evaluated explicitly without the integration limits, and the result is
\begin{align}
  \mathcal{I} & = \Omega_3^{\textrm{reg}} L^2 \frac{J^0 (J^1)^3}{2 i} \Bigg[ \sqrt{\frac{1 + i h^1/r}{1 + i h^0/r}} \left(i r (h^1 (-3 + r) + 2 i r^2) - h^0 (2 h^1 (-2 + r) + i r (1 + r)) \right) \nonumber\\
  {} & \qquad\qquad + \sqrt{\frac{1 - i h^1/r}{1 - i h^0/r}} \left(i r (- h^1 (3 + r) + 2 i r^2) - h^0 (2 h^1 (2 + r) - i r (-1 + r))\right) \Bigg]\, .
\end{align}
Taking the integration limits into account, we will consider $r \in [1,\, r_0]$ in Subsection~\ref{sec:Action}, where $r_0$ is a cutoff, i.e.,
\begin{align}
  S_{\textrm{vec}} & = \mathcal{I} (r = r_0) - \mathcal{I} (r = 1) \nonumber\\
  {} & = \Omega_3^{\textrm{reg}} L^2 \frac{J^0 (J^1)^3}{2 i} \Bigg[- 4 r_0^3 + \frac{r_0}{2} \left((h^0)^2 - 3 h^1 (4 i + h^1) - 2 h^0 (2 i + 3 h^1) \right) \nonumber\\
  {} & \qquad\qquad\qquad\qquad\quad - 2 (h^1 - i)^{3/2} (h^0 - i)^{1/2} + 6 (h^1 + i)^{3/2} (h^0 + i)^{1/2} \nonumber\\
  {} & \qquad\qquad\qquad\qquad\quad - 6 i (h^1 + i)^{1/2} (h^0 + i)^{1/2} - 2 i (h^1 + i)^{3/2} (h^0 + i)^{-1/2} \Bigg]\, .
\end{align}

Similar to the vector multiplet, for the hypermultiplet action $S_{\textrm{hyp}}$, up to the volume of the boundary manifold, the integral in the radial direction has the same integrand as the $\textrm{AdS}_4$ case, but with a different measure from $\sqrt{g}$. In the end, for the $\textrm{AdS}_4$ neutral topological black hole considered in this paper, $S_{\textrm{hyp}}$ can be expressed as
\begin{align}
  S_{\textrm{hyp}} & = - i \Omega_3^{\textrm{reg}} L^4 \int_1^{r_0} dr\, r^2 \frac{1}{r^2} \frac{g}{2 \pi G L} \left(h^0 J^0 P_0 + h^1 J^1 P_1 \right) \nonumber\\
  {} & = -i \frac{\Omega_3^{\textrm{reg}}\, g L^3}{2 \pi G} (r_0 - 1) \left(h^0 J^0 P_0 + h^1 J^1 P_1 \right) \nonumber\\
  {} & = \frac{i \Omega_3^{\textrm{reg}}\, L^2}{16 \pi G} (r_0 - 1) \left(h^0 + 3 h^1 \right)\, ,
\end{align}
where again $r_0$ is a cutoff, and we have used the attractor solutions Eq.~\eqref{eq:attSol-1}:
\begin{displaymath}
  8 g J^0 P_0 = - \frac{1}{L}\, ,\quad 8 g J^1 P_1 = - \frac{3}{L}\, .
\end{displaymath}

\bibliographystyle{utphys}
\bibliography{SRE}

\end{document}